\documentclass[12pt]{article}

\usepackage{styles/asa} % asa.sty

\usepackage{lmodern}

\usepackage[margin=1in]{geometry}
\usepackage[doublespacing, nodisplayskipstretch]{setspace} % \doublespacing, Double-spaced document text
\usepackage{parskip}
\usepackage{titlesec}
\titlespacing*{\section} {0pt}{14pt}{7pt}{ }
\titlespacing*{\subsection} {0pt}{10pt}{5pt}{ }
\titleformat{\section}  % which section command to format
  {\fontsize{15}{16}\bfseries} % format for whole line
  {\thesection} % how to show number
  {1em} % space between number and text
  {} % formatting for just the text
  [] % formatting for after the text

\titleformat*{\subsection}{\fontsize{13}{16}\bfseries}

\usepackage{float}     % floats are containers for things in a document that cannot be broken over a page, e.g. figure[htb!]
\AfterEndEnvironment{figure}{\noindent\ignorespaces}

\usepackage{amsmath}   % provide various features for displayed equations and other mathematical constructs, \begin{equation} ...
\usepackage{amssymb}   % math symbols
\usepackage{romannum}  % to generate roman numerals, \Romannum ...

{ % set spacing (stretch 1.2) for table
    \makeatletter
    \AtBeginEnvironment{tabular}{%
        \@currsize}%
    \makeatother
}

\usepackage{multirow}    % \multirow ...
\usepackage{adjustbox}   % \adjustbox ...
\usepackage{booktabs}    % create tables, \toprule, \midrule, \cmidrule, \bottomrule ...
\usepackage{makecell}    % tabular column heads and multilined cells, \makecell ...
\usepackage{threeparttable}  % footnotes for table, \tablenotes ...

\usepackage{graphicx}    % \includegraphics ...
\usepackage{caption}     % offer customization of captions in floating environments, e.g. figure and table 
\usepackage{subcaption}  % \subfigure ...
\usepackage{dashrule}  % can draw a huge variety of dashed rules, \hdashrule (horizontally dashed) ... 
\usepackage{tikz}      % to create graphic element, \tikz\draw ...

\usepackage[dvipsnames]{xcolor}    % driver-independent color extensions for LaTeX and pdfLaTeX
\usepackage{colortbl}  % allow rows and columns to be colored, and even individual cells

\usepackage{abstract}  % give control over the typesetting of the abstract environment

\renewenvironment{abstract}{
    \small
    \begin{center}
        \bfseries \abstractname\vspace{-.5em}\vspace{0pt}
    \end{center}
    \list{}{
        \setlength{\leftmargin}{0cm}%
        \setlength{\rightmargin}{\leftmargin}%
    }%
    \item\relax
}{\endlist}

\usepackage[hang]{footmisc}
\usepackage{natbib}    % \bibliographystyle, \bibliography ...
\usepackage{hyperref}  % create hyperlinks within the document
\hypersetup{
    hidelinks
}

\usepackage[toc,title,page]{appendix}

\newcommand{\bbR}{\mathbb{R}}

\newcommand{\bbE}{\mathbb{E}}  % expectation
\newcommand{\bbCov}{\mathbb{C}\textup{ov}}  % covariance

\newcommand{\bs}{\mathbf{s}}
\newcommand{\bu}{\mathbf{u}}
\newcommand{\bv}{\mathbf{v}}
\newcommand{\bx}{\mathbf{x}}

\newcommand{\bOne}{\mathbf{1}}

\newcommand{\bsbeta}{\boldsymbol{\beta}}

\newcommand{\bsphi}{\boldsymbol{\phi}}
\newcommand{\bssigma}{\boldsymbol{\sigma}}
\newcommand{\bstheta}{\boldsymbol{\theta}}

\newcommand{\cI}{\mathcal{I}}

\newcommand{\td}{\text{d}}

\newcommand{\tinhom}{\text{inhom}}
\newcommand{\thom}{\text{hom}}
\newcommand{\khat}{\widehat{K}}

\newcommand{\lamhat}{\hat{\lambda}}
\newcommand{\sumin}{\sum_{i=1}^n}
\newcommand{\bunderline}[1]{\underline{#1\mkern-4mu}\mkern4mu }
\newcommand{\paren}[1]{{\footnotesize(}#1{\footnotesize)}}

\usepackage{authblk}   % to prepare author and affiliation blocks, \author, \affil

\title{\large{\bf Addressing Duplicated Data in Spatial Point Patterns}}

\author[a]{Lingling Chen}
\author[a]{Mikyoung Jun \thanks{Corresponding author. Mikyoung Jun and Scott J. Cook acknowledge support by NSF DMS-1925119 and DMS-2105847. Mikyoung Jun also acknowledges support by NSF DMS-2413042 and Scott J. Cook acknowledges NSF DMS-2413043.}}
\author[b]{Scott J. Cook}

\affil[a]{Department of Mathematics, University of Houston}
\affil[b]{Department of Political Science, Ohio State University}

\date{ }

\begin{document}

%%%%% title
\maketitle

\vspace{-.4cm}

%%%%% abstract
%%%%% 200 or fewer words
\begin{abstract} 
\noindent Spatial point process models are widely applied to point pattern data from various applications in the social and environmental sciences. 
However, a serious hurdle in fitting point process models is the presence of duplicated points, wherein multiple observations share identical spatial coordinates. 
This often occurs because of decisions made in the geo-coding process, such as assigning representative locations (e.g., aggregate-level centroids) to observations when data producers lack exact location information. 
Because spatial point process models like the Log-Gaussian Cox Process (LGCP) assume unique locations, researchers often employ {\it ad hoc} solutions (e.g., removing duplicates or jittering) to address duplicated data before analysis.  
As an alternative, this study proposes a Modified Minimum Contrast (MMC) method that adapts the inference procedure to account for the effect of duplicates in estimation, without needing to alter the data. 
The proposed MMC method is applied to LGCP models, focusing on the inference of second-order intensity parameters, which govern the clustering structure of point patterns. 
Under a variety of simulated conditions, our results demonstrate the advantages of the proposed MMC method compared to existing {\it ad hoc} solutions. 
We then apply the MMC methods to a real-data application of conflict events in Afghanistan (2008-2009). 
\end{abstract}

\par\vfill\noindent
% 3-6 keywords that do not appear in the title

\clearpage 
\pagenumbering{arabic}

%%%%% sections
\section{Introduction}\label{sec:intro}
%%%%%
Event data (i.e., presence-only data) are commonly found in various fields, such as environmental sciences, population ecology, sociology, and political science. 
In the social sciences, for example, they are used to analyze spatial patterns of social and political behavior, including analyses of protests \citep{earl2004use}, crime \citep{krieger2015police}, terrorism \citep{lafree2019evolution}, and civil violence \citep{jcr2009}. 
These data code actions at the incident level -- that is, they provide who-did-what-when-where-and-why information on discrete events -- thereby allowing for greater insights into their determinants. 
Given continued improvements in our ability to automate event identification and extraction from secondary sources \citep{Lee2019}, event data are likely to see wider use in the future.

%%%%%
Much of the promise of event data analysis in social science applications, however, is often unrealized due to the well-known limitations of these data (e.g., missing events, geolocation errors, etc.). 
Event data in social sciences are largely drawn from media reports, which risks description bias as event details may go inaccurately reported \citep{earl2004use}. 
This can include information as basic as where the event occurred \citep{Weidmann2015}. 
Oftentimes, for example, the initial media report may contain imprecise information on the event location, providing only the district or state within which it occurred \citep{cook2022race}. 
In these instances, researchers often assign representative spatial coordinates (i.e., the observations are snapped), such as the centroid for the administrative unit in which the event was reported. 

%%%%%
While this may be the best that can be achieved, given the inherent limits on the available information, these geo-coding decisions present two problems for analysts interested in applying point process models. 
First, the assigned locations may deviate considerably from the true locations, i.e., geolocation error, threatening the validity of any inferences drawn using the observed point pattern. 
Second, the likelihood of duplicated data rises dramatically since multiple events may be assigned to the same location. Duplicated data creates issues when applying spatial point process models, as these often assume a ``simple'' process, i.e., no two points coincide in the same position \citep[][p.47]{Daley2003}. 

%%%%%
Here we consider the latter problem -- i.e., snapping-induced-duplicate data -- as it is not something that has received careful attention. While the consequences of duplicate data -- e.g., non-invertible covariance matrices -- are well known, there are not widely accepted solutions. Our survey of the literature indicates that most applied researchers adopt ad hoc remedies such as data elimination (i.e., removing duplicate observations) and data manipulation (i.e., jittering point locations). While these may evade the issue of duplicate points, this is achieved at a great cost, as these strategies induce sample selection bias and errors in location information. Recent theoretical work on temporal point patterns has instead sought to develop non-simple point process models that explicitly incorporate coincidental points \citep{Schoenberg2006}. 
However, these models are difficult to construct and estimate because non-simple processes often suffer from non-uniqueness in their conditional intensities \citep{Schoenberg2006} and are not easily extended to spatial or spatio-temporal settings. 

%%%%%
Given the limitations of these current approaches, we propose a strategy to address duplicates through adjustments in the estimation procedure, which neither requires alterations to the data nor the assumed process. 
Specifically, we propose a modified version of the Minimum Contrast (MC) estimation method by introducing a positive lower bound to the lag limit in the discrepancy measure. We demonstrate that the use of positive lower bounds can be used to address duplicates by excluding point pairs with zero-distance separation from the contrast integral. While the use of non-zero lower bounds has been considered in MC estimation elsewhere, this is typically to avoid computational instability due to large variability at near-zero distances \citep{Davies2013,Siino2018,DAngelo2023}. In extending this approach to deal with duplicate data, we also provide a practical rule for selecting the value of this lower bound based on geometric features of the sample, which should facilitate implementation of our approach in applied research involving snapping-induced duplicate data. 

%%%%%
The remainder of the paper is organized as follows. 
Section \ref{sec:motivation} introduces the causes and consequences of duplicated points and summarizes existing methods to address duplicates. 
We then provide two motivating real-world examples to demonstrate how duplicates arise and their effect on the $K$-function, a widely used summary statistic for point processes. 
In Section \ref{sec:method}, we detail our MMC estimation method for inference adjustment when analyzing observed point patterns containing duplicates. 
Then, in Section \ref{sec:simulation}, we design simulation scenarios to illustrate the performance of the proposed method as compared to existing methods. 
In Section \ref{sec:application}, we apply these same approaches to real-world conflict data from Afghanistan before concluding in Section \ref{sec:discussion}. 
Details in the choice of bandwidth for the estimation of the first-order intensity functions are provided in the Appendix. 

\section{Motivation}\label{sec:motivation}
%%%%%
In incident-level data, there are two types of duplicated data that researchers often confront: i) multiple entries for the same incident and ii) multiple incidents with the same spatial coordinates. 
In the former, multiple records of the same event are entered into the sample -- arising from multiple reports, etc. -- and should be removed in the data processing stage (i.e., de-duplication) prior to analysis. 
In the latter, duplicate data (or duplicated points) arise when separate events or incidents are assigned the same spatial coordinates. 
As these are independent events, they should each remain in the sample; however, the coincident location information can pose challenges in subsequent spatial point pattern analysis. 
This is the issue we concentrate on here. 

%%%%%
\begin{figure}[htb!]
    \centering
    \begin{tikzpicture}[scale=0.8, transform shape]
    \draw[thick] (0,0) -- (3.5,0) -- (3.5,3) -- (0,3) -- (0,0);
    \fill[red] (1.75,1.5) circle(.1);
    \foreach \p in {(0.82,1.75),(2.26,0.45),(2.88,2.65)}
    \fill \p circle(.1);
    \draw[thick,densely dashed,->] (0.82,1.75) -- (1.60,1.51); % left one 
    \draw[thick,densely dashed,->] (2.26,0.45) -- (1.82,1.35); % bottom one
    \draw[thick,densely dashed,->] (2.88,2.65) -- (1.89,1.57); % upper-right one
    \end{tikzpicture}
    \caption{Illustration of snapped event data. 
    Black dots are true locations within a district, and the red dot is the recorded location(s).}
    \label{fig:snapping}
\end{figure}

%%%%%
Duplicated points can arise in many ways; however, they typically occur due to imprecise measurement, coding choices made during data processing, or both. 
At sufficiently high levels of resolution, the probability of multiple events occurring in the same location should approach zero. 
Our measures of these incident locations, however, are rarely this precise, meaning the recorded locations nearby can and often do coincide. 
Still other times, duplicate data is produced due to deliberate decisions made in data publishing. 
When there are (geo)privacy concerns relating to the data, for example, incident locations may be ``snapped" to the nearest notable spatial feature, thereby risking duplicate points (see Figure \ref{fig:snapping}). 
In the Police.uk data \citep{Tompson2015}, for example, the actual crime event location is obfuscated by snapping to the nearest point from a predetermined list of connotatively neutral locations to protect confidentiality. 
This process -- one of the geographic masking techniques presented in \citet{Zandbergen2014} -- likely inflates the number of duplicate points we observe because each snap location may have multiple events nearby, resulting in those crimes being reassigned to the same geographic coordinates.

%%%%%
In other instances, duplicates can arise due to both data imprecision and resultant coding decisions, as in many social and political events datasets. 
To illustrate, consider the Global Terrorism Database \citep[GTD;][]{STARTdata}, which draws from publicly available information, e.g., media reports, to produce data on incidents of domestic and international terrorism. 
These media reports often lack precise information on the location of the incident, indicating only a general area in which the event took place. 
Lacking exact location information, the spatial coordinates of the event are instead assigned/snapped to the centroid of the smallest administrative region identified. 
These coding conventions are commonly found in media-sourced event data and inevitably incur duplicate points. 
In what follows, we refer to point patterns with snapping-induced duplicates as \textit{corrupted} data.

%%%%%
Regardless of their origin, duplicate points pose several challenges when applying statistical methodology to spatial point processes. As mentioned in Section \ref{sec:intro}, many point process models are based on the assumption that the underlying true processes are simple, i.e., no two points occupy the same location. Duplicated points violate this assumption, meaning that statistical procedures designed for simple point processes will be invalid or non-applicable \citep[][p.60]{Baddeley2015}. 
For example, in estimation of Log-Gaussian Cox Process (LGCP) models that involve covariance matrices, such as maximum likelihood, duplicated points lead to non-invertible covariance matrices or ill-conditioned optimization problems. 

%%%%%
Even with inference methods that do not deal with covariance matrices, such as minimum contrast estimation \citep{Diggle2003}, the presence of duplicates may produce biased estimates of the parameters that characterize the underlying spatial point process. 
Specifically, duplicated points artificially inflate the observed count of points within small distance lags (as seen in Figure \ref{fig:motivate_sigacts_ged_kinhom}). When these clusters result from the duplicates and not the underlying process, we obtain biased estimates of parameters associated with spatial clustering in the model. As such, researchers need to carefully consider the issue of duplicates in spatial point process modeling. 

%%%%%%%%%%%%%%%%%%%%%%%%%%%%%%%%%%%%%%%%%%%%%%%%%%
\subsection{Handling of Duplicates}\label{subsec:methods_duplicates}
%%%%%
There is no single, simple approach to address duplicate data fully. 
Ignoring the presence of duplicates and blindly applying inference methods is inadvisable as one would obtain unreliable and misleading results. 
However, we will later use this ``do nothing" approach as a baseline for comparison.
In our review of the relevant literature on spatial point pattern data, we found several more direct approaches to dealing with duplicates used in applied research, as discussed below. 

%%%%%
First, researchers can simply remove duplicated points from the data (Method \Romannum{1}). 
By deleting all but one observation at each location where duplication occurs \citep{Turner2009}, researchers ensure there is no duplicate data in the analysis sample. 
However, this is achieved at a substantial cost, since removing these duplicates results in a loss of information from the data and biased inferences, especially the underestimation of first-order intensity. 

%%%%%
Second, researchers may jitter the locations of duplicated points (Method \Romannum{2}). 
Here, one adds a random perturbation to the coordinates of each duplicated point \citep[e.g.,][]{Iftimi2018, Jun_Cook23}, which again effectively eliminates duplicate points. 
For example, one may jitter the $x$- and $y$-coordinates of the duplicated point independently by amounts chosen from a uniform distribution $\mathcal{U}(-d, d)$ for some $d{\,>\,}0$.  
Unlike deletion, this method allows researchers to retain all of the data; however, it is sensitive to \textit{ad hoc} choices regarding the jittering distribution(s). 
Importantly, there is no established method to guide the selection of the jittering radius, $d$. 
It needs to be large enough to effectively remove the artifacts caused by duplicated points, but small enough that it does not substantially alter features of the original spatial pattern. 
Regardless, \emph{any} jittering induces spatial measurement error in the original sample, risking biased parameter estimates.

%%%%%
Third, as with jittering, researchers may redistribute duplicated points to other randomly determined locations within some predetermined area (Method \Romannum{3}). 
For example, one may replace each duplicated point with a random point within some associated partition area, e.g., a Thiessen polygon or a geographical unit that contains the original point. 
While this is more often applied for disaggregating count data as in \citet{Shi2013}, it can also be applied to address duplicates. 
Unlike jittering in Method \Romannum{2}, this method ensures that points near the boundary of the study domain are retained. 
However, it still introduces additional noise to the data, the extent of which is subject to the size and shape of the partition areas. 
As a result, the induced error will be larger on average for events in large partitions than for events in small ones. 
Additionally, if event determinants (e.g., population) are also correlated with the size of these units, then the induced noise can produce a form of differential measurement error.

%%%%%
Finally, of course, researchers could modify their assumptions on the underlying process and modify their analysis accordingly. As noted in Section~\ref{sec:intro}, for temporal point patterns, \cite{Schoenberg2006} has proposed an approach to uniquely characterize a non-simple marked point process and transform it into a simple one. However, as it is not clear how to extend this for spatial or spatio-temporal settings, we do not consider it further here. Alternatively, researchers could choose to forgo point process modeling and simply analyze spatial aggregates, thereby evading the challenge of duplicates. By construction, this would entail a loss of information from the event-level data \citep{zhu2021promiseperilspointprocess} and possibly introduce additional threats to inference due to the change-of-support problem \citep{zhukov2024integrating}. Given this, we do not consider these alternatives further, instead focusing on approaches suited for simple spatial point processes like conflict event data.  

%%%%%%%%%%%%%%%%%%%%%%%%%%%%%%%%%%%%%%%%%%%%%%%%%%
\subsection{Political conflict data -- Afghanistan}\label{subsec:examples} 
%%%%% motivating dataset: GTD 
Political conflict events are often characterized by high heterogeneity and significant geographic clustering, as noted by \cite{Darmofal2009}. 
Capturing these spatial patterns is crucial for understanding the geographic distribution of conflicts, identifying hotspots, and uncovering underlying factors that drive political violence. However, a common challenge when working with event data is the presence of a substantial number of duplicated points.

%%%%%
To demonstrate the extent and severity of these issues in real-world data, we utilize the Global Terrorism Database (GTD). 
In the GTD, incidents of terrorism are collected from publicly available, unclassified source materials, e.g., media articles and electronic news archives \citep{STARTcodebook}. 
In addition to other event features, GTD events are assigned specific locations, with most events supplied precise latitude and longitude coordinates. 
The presumed level of accuracy of these event-locations is reported as an additional feature -- ``specificity'' -- in the data set, which is an ordinal measure ranging from 1 (very accurate) to 5 (location unknown). 
For example, $\text{specificity}{\,=\,}1$ indicates the spatial coordinates of the event are assigned at the most specific spatial resolution -- the latitude/longitude of the city, village, or town in which the attack occurred. 
Whereas, $\text{specificity}{\,=\,}2,3,4$ indicates the coordinates of the event are less accurate and assigned at the centroid of the smallest subnational administrative region identified, e.g., subdistrict, district, province, etc. Additional details are given in \cite{STARTcodebook}.

%%%%% 
\begin{figure}[htb!]
    \centering
    \begin{subfigure}{0.32\textwidth}
        \includegraphics[scale=0.49]{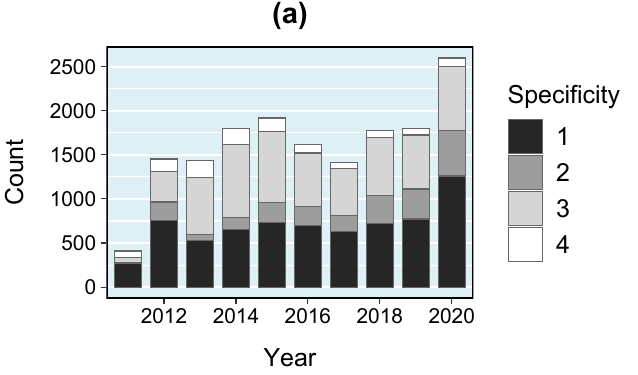}
    \end{subfigure}
    \begin{subfigure}{0.32\textwidth}
        \includegraphics[scale=0.49]{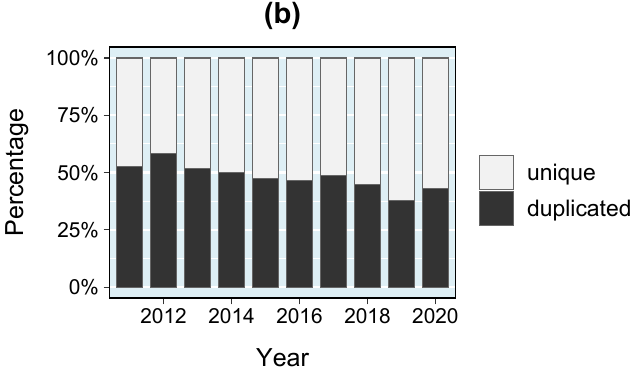}
    \end{subfigure}
    \begin{subfigure}{0.32\textwidth}
        \includegraphics[scale=0.49]{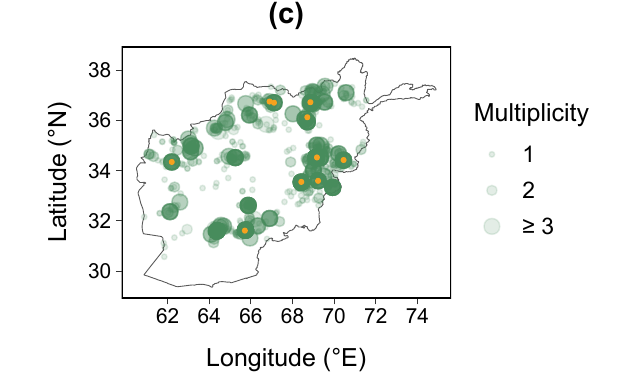}
    \end{subfigure}
    \caption{Summary of terrorism data from GTD. (a): number of terrorist attacks at each specificity level in Afghanistan from 2011 to 2020. 
    (b): the proportion of duplicates relative to the total attacks for each year.
    (c): Terrorist attacks (specificity${\,=\,}$1), in Afghanistan during 2020. 
    The top 10 most populated cities are marked with yellow. 
    Multiplicity gives the number of spatially coincident points. }
    \label{fig:motivate_gtd}
\end{figure}

%%%%%
Figure~\ref{fig:motivate_gtd} shows the prevalence of duplicates in the GTD data on Afghanistan between 2011 and 2020. 
We observe that more than half of the events have a specificity level greater than 1 (in \paren{a}), and more than $30\%$ of points in any given year are duplicated (in \paren{b}). This indicates that spatial uncertainty and duplication persist in this data set over time. 
Even among events with $\text{specificity}{\,=\,}1$, the highest level of precision, duplicate events are still quite common. So while researchers often choose to omit less precise events (ex. \citealt{python2019bayesian}) from their analysis, even this does not avoid the problem of duplicates. As shown in Figure~\ref{fig:motivate_gtd}\paren{c},  duplicates seem to be especially prevalent in urban areas, which are more likely to experience multiple attacks, yet reports may still contain insufficient information to uniquely locate them. 

%%%%% motivating dataset: SIGACTS and GED
While self-reported information on the locational uncertainty is useful, contrasting media-sourced data with more accurate (ideally gold-standard) data would obviously be preferred. 
With this in mind, we next consider a data set used in \citet{Weidmann2015}, which has drawn from two sources reporting conflict events in Afghanistan: i) the ``Significant Activities" (SIGACTS) military database, and ii) the UCDP Geo-referenced Event Dataset \citep[GED;][]{Sundberg2013}. 
For matched pairs, i.e., events appearing in both data sets, we have two reported locations -- one from the GED and one from SIGACTS -- for each observation. 
Like the GTD data, the GED data are drawn from media reports and contain many duplicates. 
On the other hand, locations in SIGACTS are measured with GPS technology; hence, they are more accurate and contain far fewer duplicates. 
The availability of conflict event data compiled independently by both the media (GED) and the military (SIGACTS) makes these data sets one of the few opportunities for a data set comparison. 

%%%%%
To illustrate the effect of corrupted data on summary statistics, we compare the estimated $K$-function \citep{Baddeley2000} from the GED to that from the SIGACTS data. Since the sample data we use contains only observations that are common to both the GED and SIGACTS data sets (i.e., matched pairs) as in \citet{Weidmann2015}, we use the estimated $K$-function of SIGACTS data, which contain few duplicates, as a reference for the estimated $K$-functions of GED data with various methods dealing with duplicates.  
Due to the clear spatial structure of the first-order intensity function of these data, we use the inhomogeneous version of the $K$-function. 
It is typically estimated by $\widehat{K}_{\tinhom}$ given later in equation \eqref{eqn:estKinhom}.

%%%%%
\begin{figure}[htb!] 
    \centering
    \includegraphics[scale=0.65]{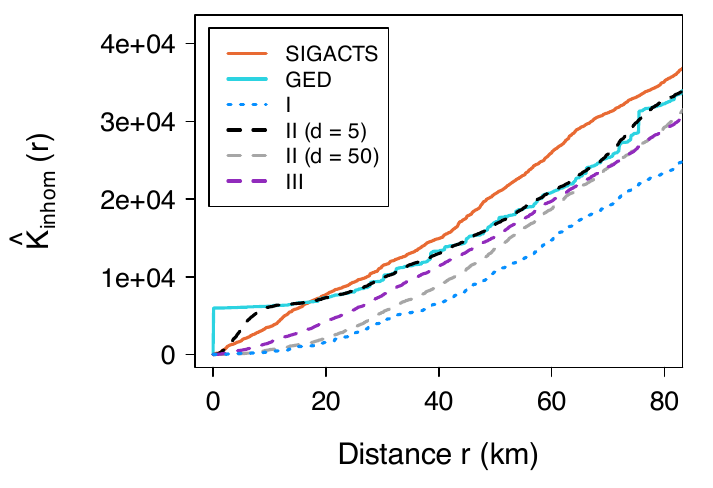}
    \caption{Estimated inhomogeneous $K$-function for SIGACTS, GED, and GED with three methods to deal with duplicates: MC-\Romannum{1}. \Romannum{2}, \Romannum{3}.  
    For MC-\Romannum{2}, the jittering amount is drawn from a uniform distribution with radius $d$ (km); for MC-\Romannum{3}, districts of Afghanistan are used as partition areas.} \label{fig:motivate_sigacts_ged_kinhom}
\end{figure}

%%%%%
Figure \ref{fig:motivate_sigacts_ged_kinhom} shows the estimated $K$-functions for the SIGACTS data, the original GED data, and the GED data with the three methods to handle duplicates described above. 
We see that the $\widehat{K}_{\tinhom}$ for the original GED data (solid blue curve) presents a large positive jump near the origin caused by duplicate observations. Moreover, after $r{\,\approx\,}20$ (km), the values are consistently lower than those for SIGACTS (solid orange curve). 
Typically, researchers would interpret this as evidence of strong clustering for small distance lags in the GED data (with duplicates); however, given the SIGACTS results (few duplicates), it seems likely to us that this result is driven by the presence of duplicates rather than the behavior of the points themselves. 

%%%%%
How does this change when employing one of the methods used to deal with duplicates? Introducing a small amount of jittering ($d{\,=\,}5$ km) for each coordinate using MC-\Romannum{2} eliminates the sharp jump at $r{\,=\,}0$, but as we move away from the origin the curve (black dashed curve) converges with the curve with the original GED data (blue curve).  Using a greater jittering radius ($d{\,=\,}50$ km) eliminates the jump near the origin, but the curve (dashed light gray curve) still does not align well with that of SIGACTS data. Similarly, relocating duplicate points randomly within the district (MC-\Romannum{3}) results in estimated $K$-function curve that is still far from the curve with SIGACTS data. Worse still is just eliminating duplicates, as in  MC-\Romannum{1}, which produces the lowest estimated $K$-function curve (blue dotted curve) due to the deleted observations. In sum, none of these approaches appears to recover the estimated $K$-function we would anticipate from a sample of events without duplicate data. This is crucial, as the estimated $K$-function curves are used to form the optimization criteria for parameter estimation, as we explore further in Section 3. 

\section{Proposed Methods}\label{sec:method}
%%%%%%%%%%%%%%%%%%%%%%%%%%%%%%%%%%%%%%%%%%%%%%%%%%
Each of the methods to deal with duplicates introduced above modifies the original spatial point pattern data in some manner, either by deleting, jittering, or redistributing observations. 
To develop a more principled method for handling duplicates, we refrain from altering data and instead adjust the inference method to accommodate duplicated data. 
We focus here on duplicate data in the context of the LGCP, a widely used model for analyzing spatial point pattern data. In the following, we first introduce some necessary preliminaries on the LGCP framework and then detail our proposed duplicate-robust inference method. While we introduce our approach in the context of an LGCP model, it can be applied to other parametric point process models, e.g. Neyman–Scott and Thomas processes, whose theoretical $K$-functions have closed-form expressions. 

%%%%%%%%%%%%%%%%%%%%%%%%%%%%%%%%%%%%%%%%%%%%%%%%%%
\subsection{Log-Gaussian Cox Process}\label{subsec:lgcp}
%%%%% LGCP
LGCPs, proposed by \cite{Moller1998}, are a flexible class of models capable of capturing complex spatial structures in point patterns. 
They have been widely used in numerous fields, such as environmental studies \citep{Valente2023}, ecology \citep{Renner2015}, and spatial epidemiology \citep{Diggle2013a}. 
% We define an LGCP model in the following way: at the first level, the process is a Poisson process conditional on the intensity function $\lambda$; at the second level, this intensity function $\lambda$ is considered a realization of the stochastic process $\Lambda$,
We define an LGCP model in the following way: at the first level, the process is a Poisson process conditional on the intensity function $\lambda$; at the second level, $\lambda$ is considered a realization of the stochastic process $\Lambda$,
\begin{equation}\label{eqn:LGCP}
    \log\Lambda(\mathbf{s}) = m(\bs) + Z(\bs), 
\end{equation}  
where $m(\bs){\,=\,}{\alpha + \bv(\bs)^{T}\bsbeta}$, $m(\bs)$ is the deterministic mean structure, $\bv(\bs)$ is a vector of covariates at location $\bs{\,\in\,}\bbR^2$, $\alpha$ and $\bsbeta$ are the associated parameters to be estimated. The term $Z{\,=\,}\{Z(\bs) {:\,} {\bs\in\bbR^2}\}$ represents a Gaussian random field (GRF), capturing the ``residual" correlation.
Throughout the remainder of the paper, we assume that $Z$ has a constant mean $\bbE[Z(\bs)]{\,=\,}-\sigma^2/2$, and an isotropic exponential covariance function
\begin{equation}\label{eqn:covariance}
    c(h; \phi,\sigma^2) = \bbCov\{Z(\bs), Z(\bu)\} = \sigma^2\exp(-h/\phi), \quad h=\|\bs-\bu\|, 
\end{equation}
where $\sigma^2$ is the variance parameter, and $\phi$ is the spatial range parameter.

%%%%%
Three prevalent statistical inference methods for LGCP models are maximum likelihood inference, Bayesian inference, and MC estimation. 
Maximum likelihood inference and Bayesian inference, while powerful, may be challenging due to the intractability of the likelihood functions \citep{Baddeley2007, Moller2017} even when duplicates are not present. 
When duplicates are present, these methods result in non-invertible covariance matrices (for log-intensity of LGCP) due to the fact that some off-diagonal elements of covariance matrices are equal to the variance. 
MC estimation, in contrast, does not rely on covariance matrices and can be applied directly to LGCPs with duplicates, though they still suffer from the influence of duplicates.
Hence, this study endeavors to adapt the MC method to accommodate the impact of duplicates within the LGCP framework.

%%%%%%%%%%%%%%%%%%%%%%%%%%%%%%%%%%%%%%%%%%%%%%%%%%
\subsection{Modified Minimum Contrast}
%%%%% 
The core idea of the MC method is to identify parameters that minimize the discrepancy between the theoretical summary descriptor, which involves the unknown parameters, and its empirical counterpart \citep{Moller1998, Diggle2003}. 
Here, we use the $K$-function as the summary descriptor that characterizes the second-order properties of a point process. 
Let $K(\cdot;\bstheta)$ denote the theoretical $K$-function and $\khat$ the corresponding empirical estimate, then a typical measure of discrepancy \citep{Guan2007} is 
\begin{equation}\label{eqn:U_general}
    U(\bstheta) = \int_0^{r_{\max}} \left\{ \widehat{K}(r)^{0.25} - K(r;\bstheta)^{0.25} \right\}^2\,\td r, 
\end{equation}
where $r_{\max}$ is a given positive constant.  
\cite{Diggle2003} suggests that a good rule of thumb is to set $r_{\max}=\frac{1}{4}\min(L_x, L_y)$ where $L_x$ and $L_y$ represent the maximum width and height of the study domain. 
However, in the case of heterogeneous point patterns, to capture the local features, a smaller $r_{\max}$ should be used as in \citet{Davies2013}. 

%%%%%
The mathematical form of $K(\cdot;\bstheta)$ is known for many point processes, including the LGCP defined by a GRF with an isotropic exponential covariance function (as in equation \eqref{eqn:covariance}). 
Substituting its parametric $K$-function in to \eqref{eqn:U_general} the discrepancy measures becomes
\begin{equation}\label{eqn:U_lgcp}
    U(\phi, \sigma^2) = \int_0^{r_{\max}} \bigg[ \khat(r)^{0.25}-\Big\{2\pi\int_0^r s\cdot\exp\big( \sigma^2 e^{-s/\phi} \big)\td s\Big\}^{0.25}\,\bigg]^2\td r. 
\end{equation}
We refer to the minimizer of $U(\phi, \sigma^2)$ as the MC estimator

%%%%%
Calculating the empirical $K$-function proceeds in the standard fashion. 
With a homogeneous LGCP, $\widehat{K}_{\thom}(r)$ is given by
\begin{equation}\label{eqn:estKhom}
    \widehat{K}_{\thom}(r) := \frac{1}{\lamhat}\, \sum_{i,\,j\,\in\,\cI_W}^{\neq} \frac{\bOne\big[\, \|\bs_i-\bs_j \| \leq r \,\big]}{N}\cdot e\left( \bs_i,\bs_j \right), \qquad r\geq 0, 
\end{equation}
where $N$ is the number of observed points inside a bounded window $W\subset \bbR^2$, $\cI_W$ is the index set of the $N$ events in $W$, $e(\bs,\bu)$ is an edge correction factor, and $\lamhat {\,=\,} \frac{N-1}{|W|}$ is the estimated intensity. 
Following \citet{Baddeley2000}, when the LGCP is inhomogeneous the $K$-function is instead estimated by 
\begin{equation} \label{eqn:estKinhom}
    \widehat{K}_{\tinhom}(r) := \sum_{i,\,j\,\in\,\cI_W}^{\neq} \frac{\bOne\big[\, \|\bs_i-\bs_j \| \leq r \,\big]}{\lamhat(\bs_i)\,\lamhat(\bs_j)\,|W|}\cdot e(\bs_i,\bs_j), \qquad r\geq0,
\end{equation}
where $\lamhat(\cdot)$ is an estimate of the intensity function $\lambda$ often by kernel smoothing approaches \citep{Davies2018, Davies2018a}. 
We defer the details of estimating $\lambda$ for both homogeneous and inhomogeneous cases to the Appendix.

%%%%%
When duplicates are present in data, as we observed in Figure \ref{fig:motivate_sigacts_ged_kinhom}, $\widehat{K}$ overestimates $K(r)$ when $r$ is small. 
In other words, the expected number of extra points within a relatively short distance of a randomly chosen point is overestimated, i.e., $\widehat{K}(0) > 0 = K(0)$.
Then, since we know that there will be an erroneous discrepancy between $\widehat{K}(r)$ and $K(r)$ at small values of $r$, it is natural to truncate the integral to reflect only the true discrepancies.
Hence, we modify the discrepancy measure in equation (\ref{eqn:U_lgcp}) by adding a tuning parameter $\delta>0$ as the lower limit of the integral, that is  
\begin{equation}\label{eqn:U_d}
    U_{\delta}(\phi, \sigma^2) = \int_\delta^{r_{\max}}\bigg[ \khat(r)^{0.25}-\Big\{2\pi\int_0^r s\cdot\exp\big( \sigma^2 e^{-s/\phi} \big)\td s\Big\}^{0.25}\,\bigg]^2\td r. 
\end{equation}
The resulting minimizer we call the modified minimum contrast (MMC) estimator.
In choosing $\delta$, our belief is that its optimal value is associated with the geometry of the partitions for the study region, say the length of the cell on a regular grid. 
This will be discussed in greater detail in Section \ref{subsec:tuning_pars}.

%%%%%
Our approach deviates from convention with the $K$-function in MC estimation, where common practice is to set the lower bound at zero \citep{Diggle2003, Waagepetersen2007, Davies2013, Zhu2025}. However, prior work on MC estimation has discussed the use of a positive lower bound when using the $g$-function (pair correlation function) as the summary descriptor \citep{Davies2013, Siino2018, DAngelo2023, Moller2003}. There, the need for a positive lower bound arises because the $g$-function cannot be evaluated at zero: its estimator involves a $1/r$ term, making it undefined at distance $r{\,=\,}0$ \citep[][p.229]{Baddeley2015}. To avoid resulting computational instability, researchers are advised to supply a positive lower bound. The particular choice is subjective, though often researchers use the lowest interpoint distance.

%%%%%
Here we show that a non-zero lower bound can have benefits beyond numerical stability. By assigning a positive lower bound we resolve the issue of duplicates, as point pairs with zero-distance separation are excluded from the contrast integral. To achieve this, however, we need to select a lower limit. Obviously, we cannot employ the decision-rule used in prior $g$-function work, since the lowest interpoint distance is zero due to the duplicates themselves. Instead we propose that researchers exploit geometric features of the sample data in making these choices. As noted above, \cite{Diggle2003} suggests a rule of thumb for selecting $r_{max}$ based the extent of the spatial sample, i.e. one-quarter of the length of the study domain. Similarly, we believe that the choice of the lower bound can be guided by geometric features of the sample, namely, the support of the assumed snapped data lattice (i.e., the scale). As discussed in Section~\ref{sec:motivation} researchers often have information on the corruption process, and in Section~\ref{subsec:tuning_pars} we demonstrate how this information can be used to guide the selection of the lower bound. 

\section{Simulation Study}\label{sec:simulation}
%%%%% 
To evaluate the finite-sample performance of our proposed MMC method against commonly used alternatives for fitting LGCP models, we conduct a series of simulation studies under different levels of duplication. 
As mentioned in Section \ref{sec:intro}, we assume that the underlying process is simple across all scenarios, simulating point patterns from an LGCP with varying sample geometries, mean structures, and spatial correlations (specific parameter settings are summarized in Table \ref{tab:sim_scenarios}). Importantly for our study, we then contaminate these realizations by introducing duplicates through a corruption mechanism (detailed below), which parallels the real-world data management practices described in Section \ref{sec:motivation}. To benchmark estimator performance, we use both the true parameters and the MC estimates obtained from uncorrupted simulations, comparing these with estimates obtained from analyzing the corrupted data. This design allows us to evaluate how well each approach mitigates the impact of duplicates in recovering the clustering (i.e., second-order) characteristics of the process. 

%%%%% table 1: simulation study settings
\begin{table}[ht]
\caption{Simulation study settings}
\label{tab:sim_scenarios}
\centering
\adjustbox{width=6.2in}{
\begin{tabular}{cccccr}
\toprule
\textbf{LGCP} & \textbf{Label} & \textbf{Domain} & \multicolumn{1}{c}{$\mathbf{m}(\bs)$} & $(\bsphi,\,\bssigma^2)$\\ 
\midrule
\textbf{Homogeneous} & H.1 - H.3 & $[0,\,810]^2$ & constant: $=\log(\frac{\bbE(N)}{|W|})=\log(\frac{1000}{810^2})$ & \multirow{4}{*}{\makecell[c]{$1{:\,(15, 2)\,}$\\ $2{:\,(20, 2)\,}$\\ $3{:\,(30, 2)\,}$}}\\ 
\cmidrule(lr){1-4}
\multirow{2}{*}{\textbf{Inhomogeneous}} & IH1.1 - IH1.3 & $[0,\,810]^2$ & $-7.0753 -0.0018\,s_1 +0.0026\,s_2$ & \\
\cmidrule(lr){2-4}
 & IH2.1 - IH2.3 & Afghanistan & $-1.16 -0.392\,\text{log\_cities} -1.075\,\text{log\_ring}$ & \\
\bottomrule
\end{tabular}
}
\begin{tablenotes}
\scriptsize
\item \hspace{0.15cm} $\mathbf{s}=$ ($s_1$, $s_2$) is the location vector, with $s_1$ and $s_2$ the x and y coordinates respectively. 
\item \hspace{0.15cm} \textit{log\_cities} represents the log of distance to the top 10 most populated cities in Afghanistan.
\item \hspace{0.15cm} \textit{log\_ring} represents the log of distance to the ring road in Afghanistan. 
\item \hspace{0.15cm} The unit of distance for IH2.1-IH2.3 is km. 
\end{tablenotes}
\end{table}

%%%%%
As described in Section \ref{subsec:lgcp}, the intensity function \eqref{eqn:LGCP} in the LGCP model is a log-linear function of a deterministic mean structure, $m(\bs)$, and a Gaussian random field $Z(\bs)$ which captures spatial dependence.
Here, $Z(\bs)$ is assumed to have a mean $\bbE\{Z(\bs)\}{\,=\,}{-\,}\sigma^2/2$ and an exponential covariance function $\bbCov\{Z(\bs), Z(\bu)\}{\,=\,}\sigma^2\exp(-\|\bs-\bu\|/\phi)$. 
As detailed in Table \ref{tab:sim_scenarios}, we design nine distinct scenarios categorized by varying levels of inhomogeneity and spatial dependence.
Specifically, we consider three different mean structures of the log-intensity: i) a constant mean, i.e., a homogeneous structure (H), ii)  a spatially varying mean (i.e., an inhomogenous structure), which is linear function of the location coordinates (IH1), iii) a spatially varying mean that is a linear combination of spatially heterogeneous covariates (IH2).  
Under each mean scenario, we also vary the spatial correlation, $\phi$, in the second-order intensity structure $(\phi, \sigma^2)$ with values of (15,2), (20,2), and (30,2) considered.

\begin{figure}[htb!]
    \centering
    \begin{subfigure}{0.32\textwidth}\centering
        \includegraphics[scale=0.8]{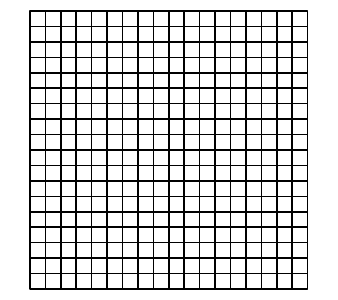}
    \end{subfigure}
    \centering
    \begin{subfigure}{0.32\textwidth}\centering
        \includegraphics[scale=0.8]{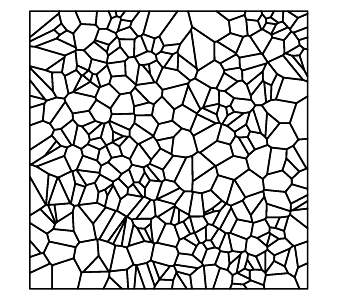}
    \end{subfigure}
    \begin{subfigure}{0.32\textwidth}\centering
        \includegraphics[scale=0.9]{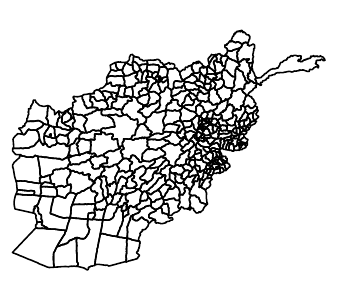}
    \end{subfigure}
    \caption{(Left) an $18\times 18\,(=324)$ regular grid with individual cell area $45^2$, (middle) a Dirichlet tessellation with 324 partitions, and (right) 328 districts of Afghanistan.}
    \label{fig:sim_admin}
\end{figure}

%%%%%
To facilitate comparison across these scenarios and increase the correspondence with our main application, we use a $[0,810]\times[0,810]$ square as the domain in H and IH1. 
This has an area roughly equal to that of Afghanistan itself (approx. 652,861 km\textsuperscript{2}), which is used as the domain in IH2. 
In each scenario, we simulate 1,000 point patterns from the LGCP, which we refer to as the ``uncorrupted'' data. 
We control the size of the simulated point pattern by adjusting the coefficients in $m(\bs)$ to ensure that the expected number of points $\bbE(N)$ is approximately 1,000.
To introduce duplicates in our simulated data, we then ``corrupt" the point pattern by selecting a certain percentage of points randomly and relocating these points to the centroid of their respective partitions. 
We apply three levels of corruption (20\%, 40\%, and 60\%) across different partitioning schemes, including regular grids, Dirichlet tessellations, and administrative district boundaries (Figure \ref{fig:sim_admin}).
This allows us to evaluate how the geometry of the corruption process affects second-order parameter estimation.

%%%%% fig: admin divisions

%%%%%%%%%%%%%%%%%%%%%%%%%%%%%%%%%%%%%%%%%%%%%%%%%%
\subsection{Results}\label{subsec:sim_res}
%%%%
In this section, we evaluate the performance of our proposed MMC approach against existing methods to recover accurate second-order parameters given duplicate data.
Based on the criteria for choosing $\delta$ (discussed later in Section \ref{subsec:tuning_pars}), we fix $\delta{\,=\,}17$ in MMC.

\begin{figure}[H]
    \centering
    \begin{subfigure}{0.42\textwidth}
    \caption{H.1 (regular grid): $(\phi=15, \sigma^2=2)$}
        \includegraphics[scale=0.55]{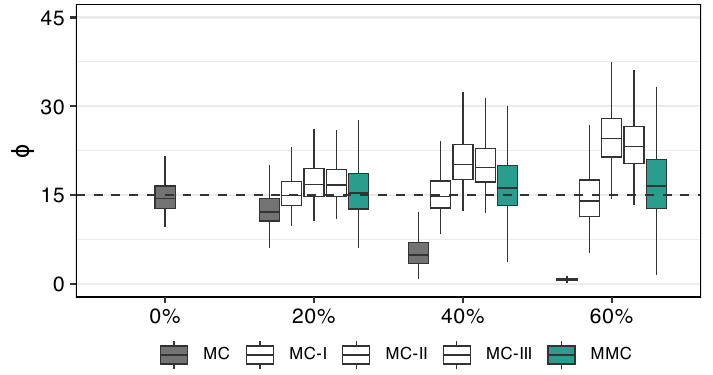}
    \end{subfigure}
    \begin{subfigure}{0.42\textwidth}
        \includegraphics[scale=0.55]{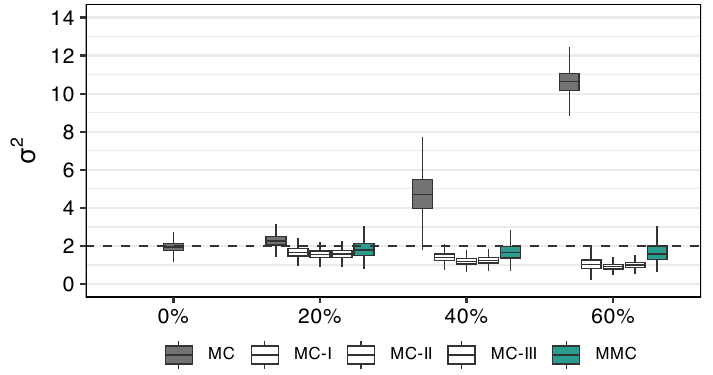}
    \end{subfigure}
    \begin{subfigure}{0.42\textwidth}
    \caption{H.2 (regular grid): $(\phi=20, \sigma^2=2)$}
        \includegraphics[scale=0.55]{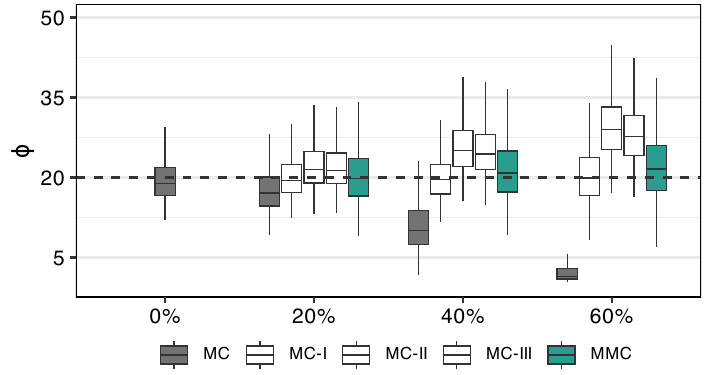}
    \end{subfigure}
    \begin{subfigure}{0.42\textwidth}
        \includegraphics[scale=0.55]{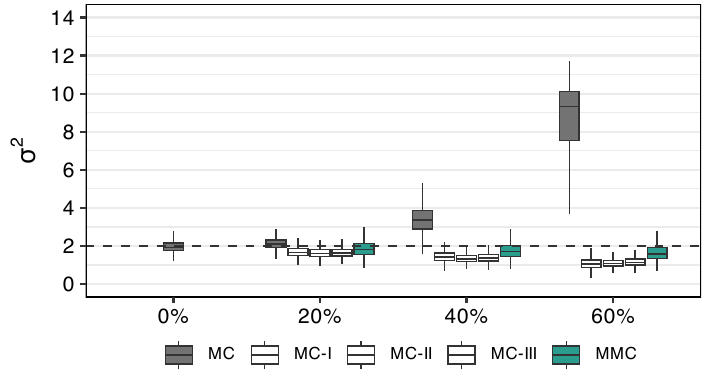}
    \end{subfigure}
    \begin{subfigure}{0.42\textwidth}
    \caption{H.3 (regular grid): $(\phi=30, \sigma^2=2)$}
        \includegraphics[scale=0.55]{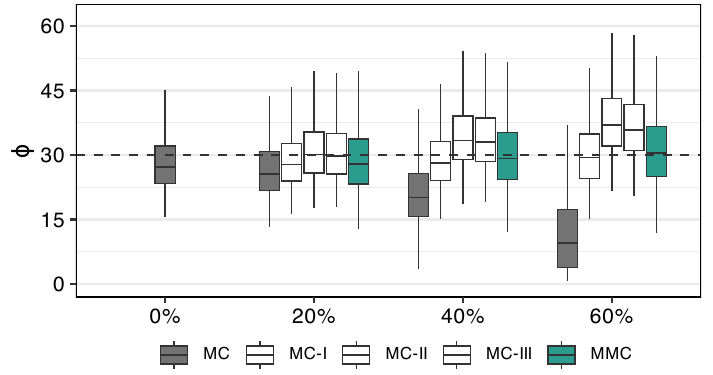}
    \end{subfigure}
    \begin{subfigure}{0.42\textwidth}
        \includegraphics[scale=0.55]{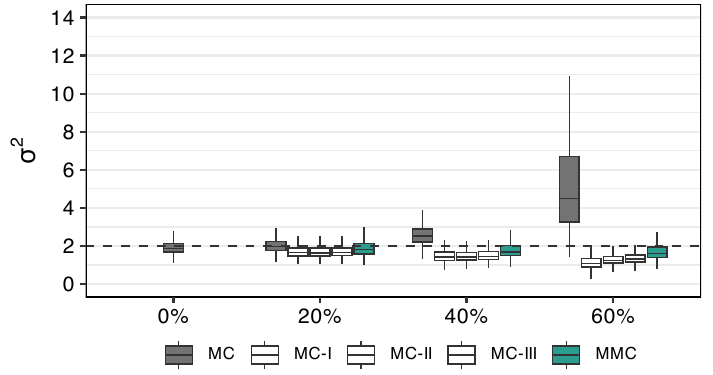}
    \end{subfigure}
    \caption{Estimates of ($\phi,\sigma^2$) for homogeneous cases across different levels of corruption on a regular grid (based on 1,000 simulations). 
    The jittering radius in MC-\Romannum{2} is 25.
    For MMC, $\delta=17$.
    The dashed lines mark the true parameter values. 
    Outliers are not plotted for legibility.} 
    \label{fig:corruptLevels_hr}
\end{figure}

%%%%
We first consider homogeneous LGCP cases (H.1–H.3) with an expected 1,000 points per trial. 
Note that the $K$-function is estimated by \eqref{eqn:estKhom}, and $\lambda$ is constant in this case. 
Figure \ref{fig:corruptLevels_hr} reports a detailed comparison of the different methods applied to homogeneous cases under various levels of data corruption on a regular grid.
Here, MC refers to the usual MC (``do nothing"), MC-I refers to Method I, MC-II refers to Method II, and MC-III refers to Method III, as described in Section~\ref{subsec:methods_duplicates}. 
The dashed lines indicate the true parameter values.
The MC estimates for simulated point patterns with no corruption (0\%) serve as a reference, which reflects how well the MC method can recover the true parameters when our data has no contamination. 

%%%%% (H) fig: different levels of corrupted data (regular grid)

\begin{figure}[H]
    \centering
    \begin{subfigure}{0.42\textwidth}
    \caption{H.1 (tessellation): $(\phi=15, \sigma^2=2)$}
        \includegraphics[scale=0.55]{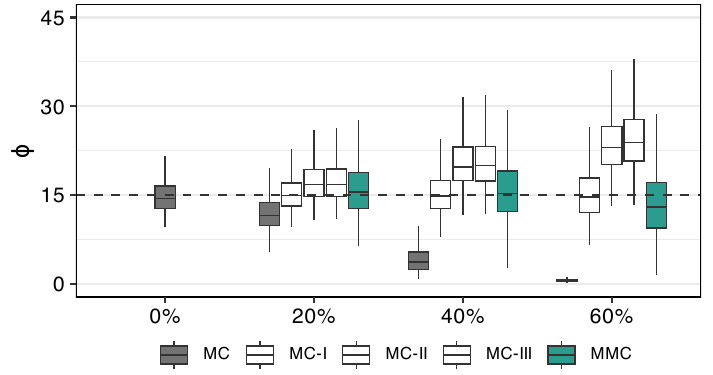}
    \end{subfigure}
    \begin{subfigure}{0.42\textwidth}
        \includegraphics[scale=0.55]{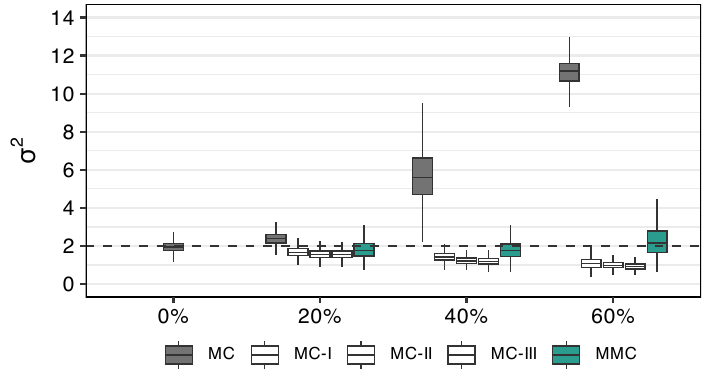}
    \end{subfigure}
    \begin{subfigure}{0.42\textwidth}
    \caption{H.2 (tessellation): $(\phi=20, \sigma^2=2)$}
        \includegraphics[scale=0.55]{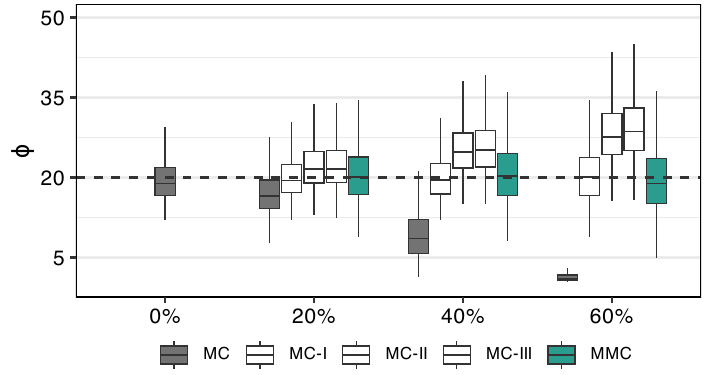}
    \end{subfigure}
    \begin{subfigure}{0.42\textwidth}
        \includegraphics[scale=0.55]{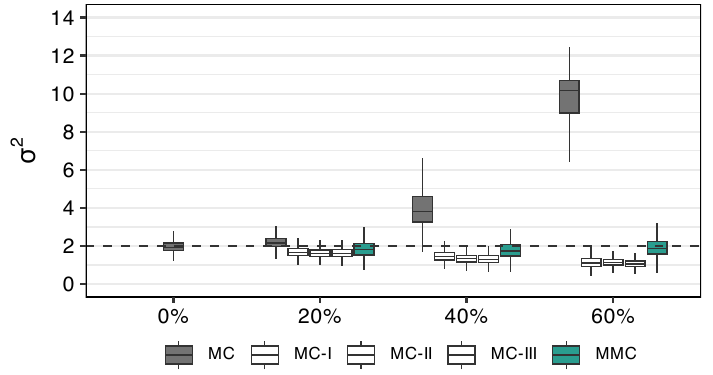}
    \end{subfigure}
    \begin{subfigure}{0.42\textwidth}
    \caption{H.3 (tessellation): $(\phi=30, \sigma^2=2)$}
        \includegraphics[scale=0.55]{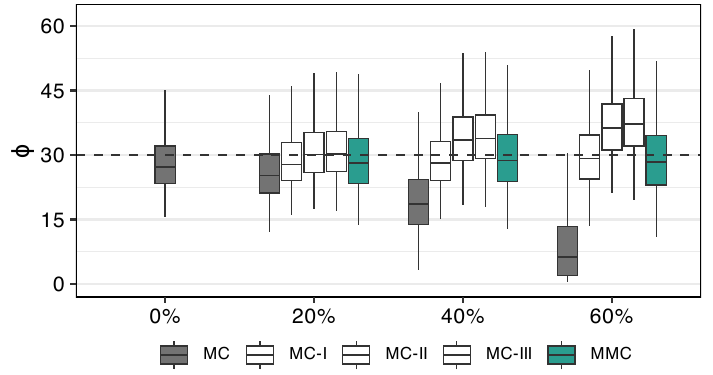}
    \end{subfigure}
    \begin{subfigure}{0.42\textwidth}
        \includegraphics[scale=0.55]{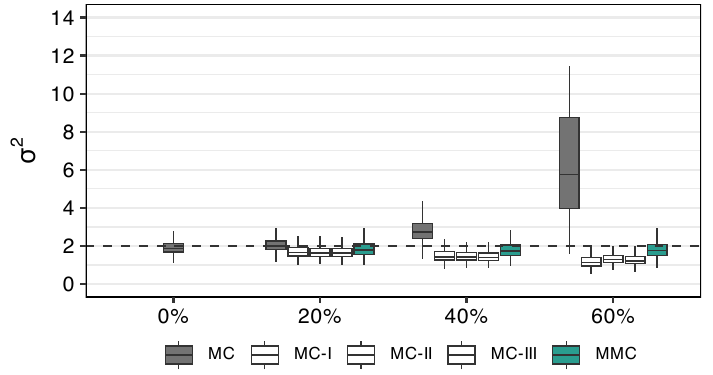}
    \end{subfigure}
    \caption{Same as Figure~\ref{fig:corruptLevels_hr}, except that a tessellation grid is used. } % $\nutol=0.03$ for MCwN.
    \label{fig:corruptLevels_ht}
\end{figure}

%%%%%
We observe that, at low levels of corruption (20\%), all methods perform reasonably well. 
However, as corruption increases, the standard MC method shows increasing bias. 
% MC-I (deletion) performs reasonably well in estimating $\phi$ across all corruption levels while considerably underestimating $\sigma^2$.
MC-I preserves the estimate of $\phi$ but generally underestimates $\sigma^2$, especially for data with larger proportion of duplicates, leading to incorrect inferences about spatial variation. 
% MC-II (jittering) and MC-III (redistribution) show similar patterns of performance. 
% Both methods tend to overestimate $\phi$ and underestimate $\sigma^2$ with a more pronounced bias at 60\% corruption. 
MC-II and MC-III both tend to overestimate $\phi$ and underestimate $\sigma^2$, with the bias becoming more pronounced at 60\% corruption.
Note that method MC-II requires a nontrivial choice to be made regarding the jittering radius. 
As discussed in Section \ref{sec:motivation}, there is no standard approach to guide this choice.
Our choice of jittering radius was $25$, producing neighborhoods slightly larger than one grid cell. 
In contrast, MMC performs comparably well in capturing both $\phi$ and $\sigma^2$ across all levels of corruption, successfully mitigating the effects of duplication. 
Similar findings hold for the tessellation-based corruption scenario (Figure \ref{fig:corruptLevels_ht}).
The MMC method consistently outperforms the alternatives.

%%%%% (H) fig: different levels of corrupted data (tessellation)

\begin{figure}[htb!]
    \centering
    \begin{subfigure}{0.42\textwidth}
    \caption{IH1.1 (regular grid): $(\phi=15, \sigma^2=2)$}
        \includegraphics[scale=0.55]{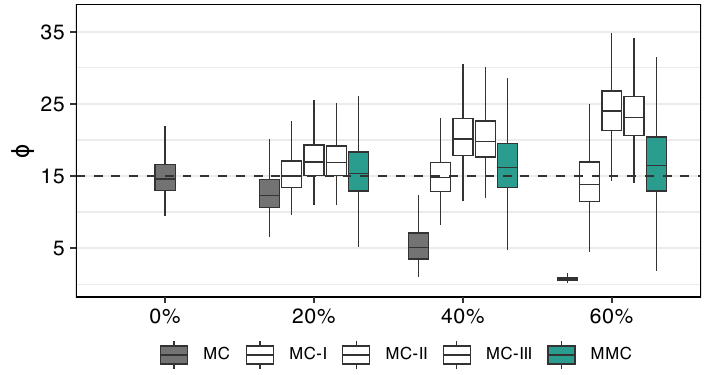}
    \end{subfigure}
    \begin{subfigure}{0.42\textwidth}
        \includegraphics[scale=0.55]{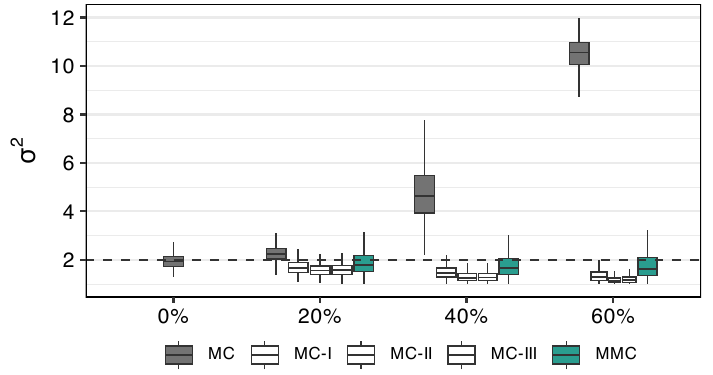}
    \end{subfigure}
    \begin{subfigure}{0.42\textwidth}
    \caption{IH1.2 (regular grid): $(\phi=20, \sigma^2=2)$}
        \includegraphics[scale=0.55]{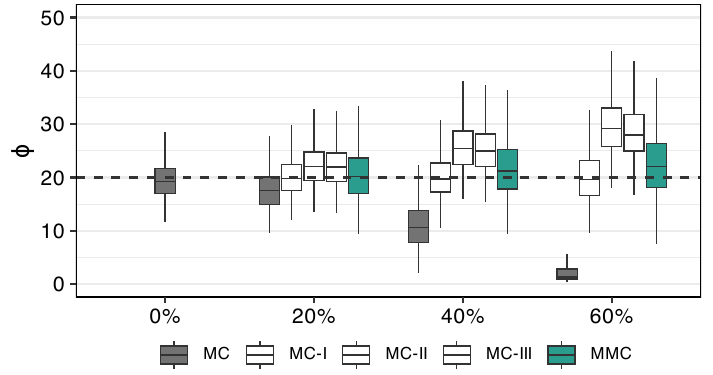}
    \end{subfigure}
    \begin{subfigure}{0.42\textwidth}
        \includegraphics[scale=0.55]{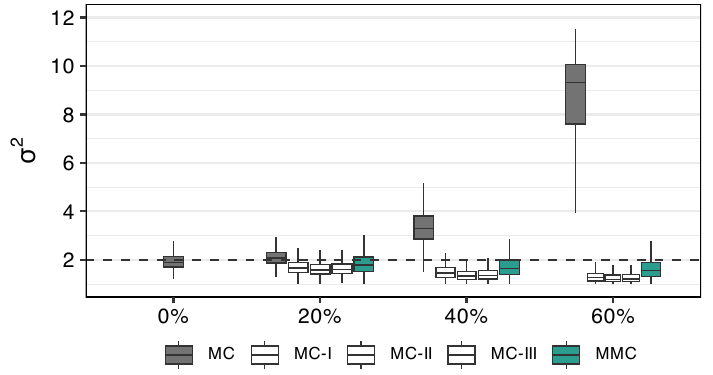}
    \end{subfigure}
    \begin{subfigure}{0.42\textwidth}
    \caption{IH1.3 (regular grid): $(\phi=30, \sigma^2=2)$}
        \includegraphics[scale=0.55]{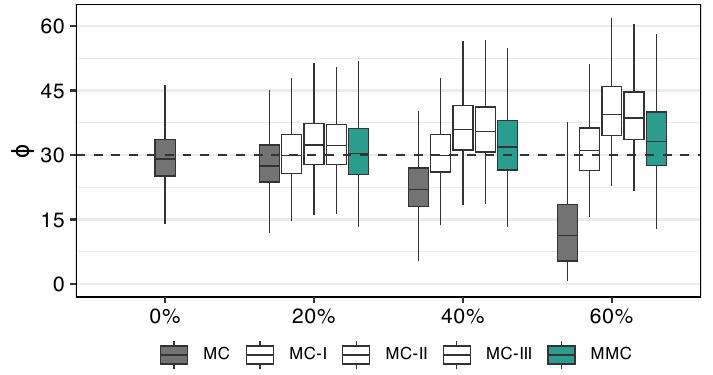}
    \end{subfigure}
    \begin{subfigure}{0.42\textwidth}
        \includegraphics[scale=0.55]{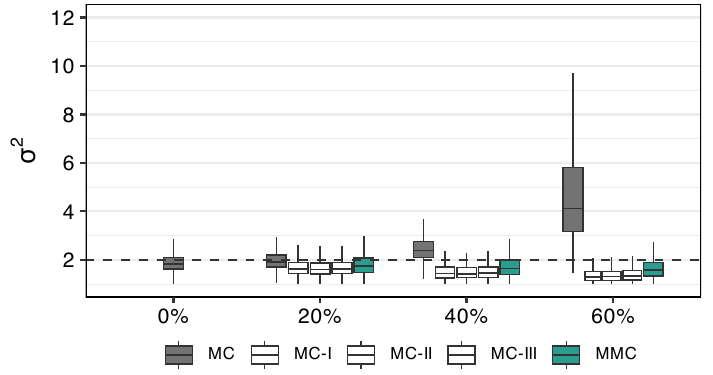}
    \end{subfigure}
    \caption{Same as Figure~\ref{fig:corruptLevels_hr}, except that this is for  inhomogeneous cases (IH1). }
    \label{fig:corruptLevels_ih1r}
\end{figure}

\begin{figure}[htb!]
    \centering
    \begin{subfigure}{0.42\textwidth}
    \caption{IH1.1 (tessellation): $(\phi=15, \sigma^2=2)$}
        \includegraphics[scale=0.55]{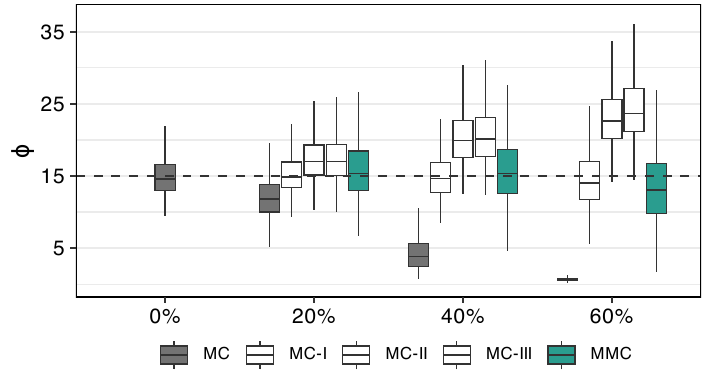}
    \end{subfigure}
    \begin{subfigure}{0.42\textwidth}
        \includegraphics[scale=0.55]{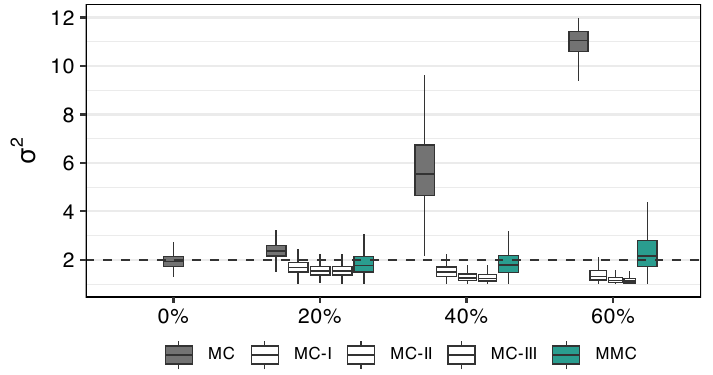}
    \end{subfigure}
    \begin{subfigure}{0.42\textwidth}
    \caption{IH1.2 (tessellation): $(\phi=20, \sigma^2=2)$}
        \includegraphics[scale=0.55]{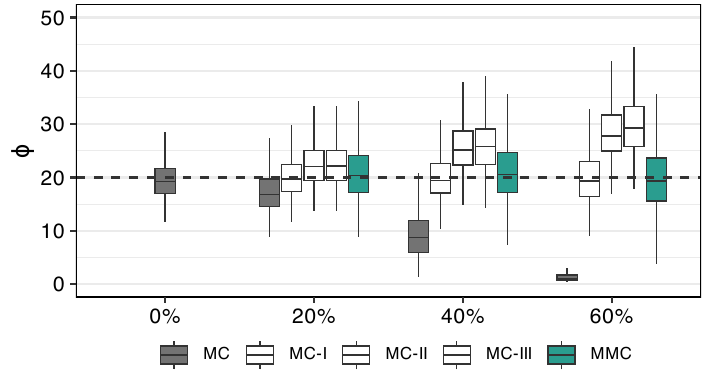}
    \end{subfigure}
    \begin{subfigure}{0.42\textwidth}
        \includegraphics[scale=0.55]{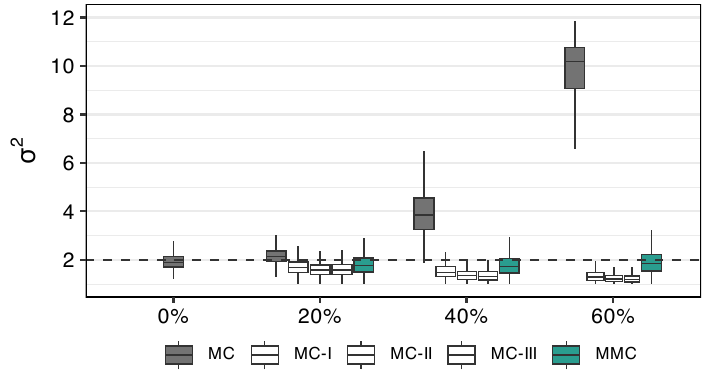}
    \end{subfigure}
    \begin{subfigure}{0.42\textwidth}
    \caption{IH1.3 (tessellation): $(\phi=30, \sigma^2=2)$}
        \includegraphics[scale=0.55]{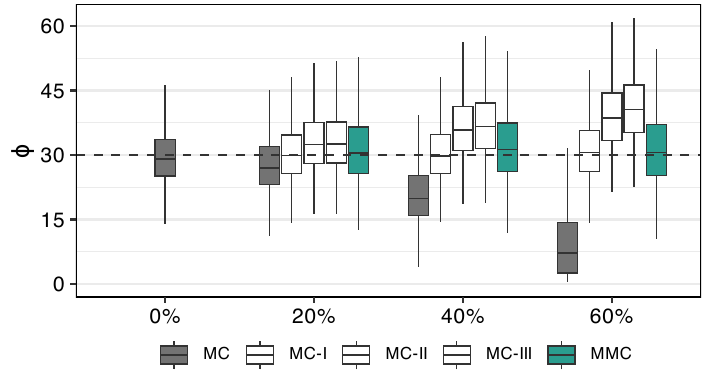}
    \end{subfigure}
    \begin{subfigure}{0.42\textwidth}
        \includegraphics[scale=0.55]{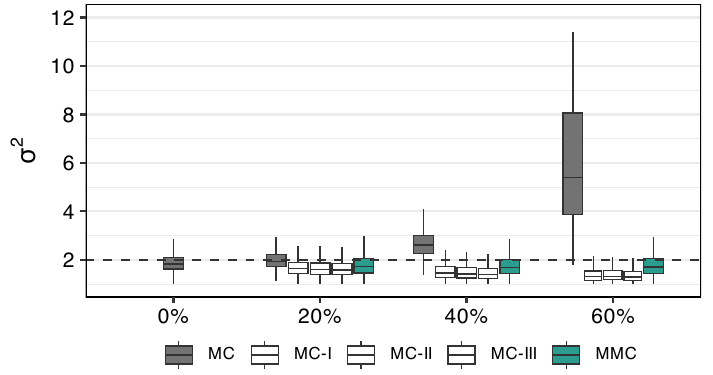}
    \end{subfigure}
    \caption{Same as Figure~\ref{fig:corruptLevels_ih1r}, except that a tessellation grid is used. } 
    \label{fig:corruptLevels_ih1t}
\end{figure}

\begin{figure}[htb!]
    % \centering
    \hspace{0.68cm}
    \begin{subfigure}[b]{\textwidth}
        \includegraphics[scale=0.64]{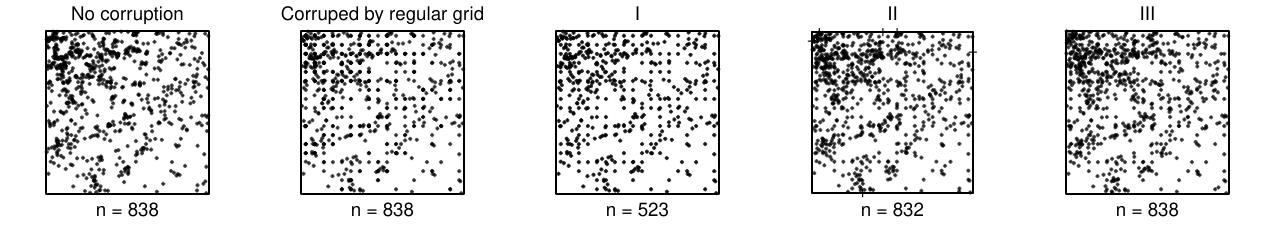} 
    \end{subfigure}
    \vfill\hspace{0.68cm}
    \begin{subfigure}[b]{\textwidth}
        \includegraphics[scale=0.75]{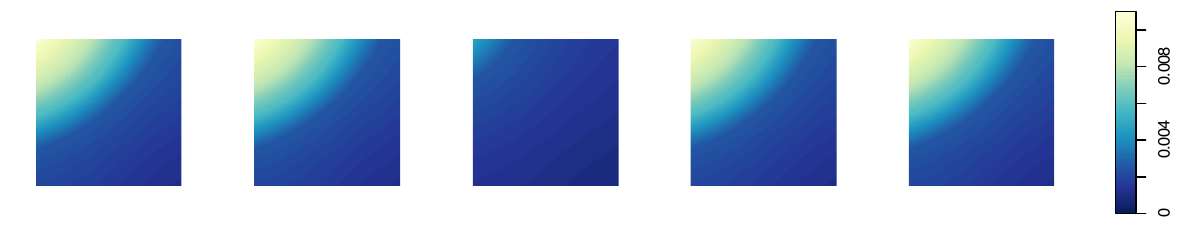}
    \end{subfigure}
    \vfill\vspace{-0.25cm}
    \caption{Top: a single realization for IH1.1 ($\phi=15$) with no corruption (left), corrupted data (60\%) by regular grid (second from the left), and the corrupted data with Methods \Romannum{1}, \Romannum{2} and \Romannum{3} to deal with duplicates. 
    Bottom: estimated intensity by the fixed-bandwidth kernel smoothing ($h=270$) for each case. 
    }
    \label{fig:ih1_inten}
\end{figure}

%%%%%
% We next consider inhomogeneous cases (IH1.1-IH1.3), with a smoothly varying intensity surface (a linear function of coordinate components). 
Next, we consider inhomogeneous LGCP cases. 
As in \eqref{eqn:estKinhom}, this case involves the estimation of $\lambda$, which requires the choice of bandwidth for kernels. 
We refer to the Appendix for details on the choice of bandwidths for the IH1 and IH2 cases.  
For the IH1, we set $h{\,=\,}270$, 285, and 325 (see the equation \eqref{eqn:inhom_lamEst1} in the Appendix), for $\phi{\,=\,}15$, 20, and 30, respectively. 
Results are given in Figures~\ref{fig:corruptLevels_ih1r} and \ref{fig:corruptLevels_ih1t}. 
In this case, the findings are similar to the homogeneous case: MMC returns estimates closer to the truth than either MC-\Romannum{2} or MC-\Romannum{3}, and as good if not better than MC-\Romannum{1} (depending on the scenario). 
While MC-\Romannum{1} seems to capture second-order properties fairly well (especially in terms of $\phi$), this comes at the expense of accurately capturing the first-order intensity. 
As shown in Figure \ref{fig:ih1_inten}, deleting the duplicate data causes the model to underestimate the first-order intensity structure. 
This has obvious consequences in applied research where the first-order intensity, often expressed as a function of spatial covariates, is central. 

%%%%% (IH1) fig: different levels of corrupted data (regular grid)

%%%%% (IH1) fig: different levels of corrupted data (tessellation)

%%%%%

%%%%%
We now discuss the other inhomogeneous case with a more realistic first-order intensity structure (IH2). 
As detailed in Table \ref{tab:sim_scenarios}, for IH2, we design a first-order intensity function as a linear combination of two covariates to generate point patterns with moderate-to-high heterogeneity.
Specifically, this function has a strong negative association with the log of distance to the ring road \citep{afgRingRoad} and a relatively weaker negative association with the log of distance to the top ten most populated cities in Afghanistan \citep{afgCities10}.
In other words, the intensity is greater near highly populated areas. 
For details on the bandwidth selection, we refer to the Appendix. 
The global bandwidths $h_0$ (see the equation \eqref{eqn:inhom_lamEst2} in the Appendix) are set to be 118, 119, and 121, for IH2.1, IH2.2, and IH2.3, respectively. 

%%%%% (IH2) fig: different levels of (randomly) corrupted data, h_0 = {118, 119, 121}
\begin{figure}[H]
    \centering
    \begin{subfigure}{0.42\textwidth}
    \caption{IH2.1 (districts): $(\phi=15, \sigma^2=2)$}
        \includegraphics[scale=0.55]{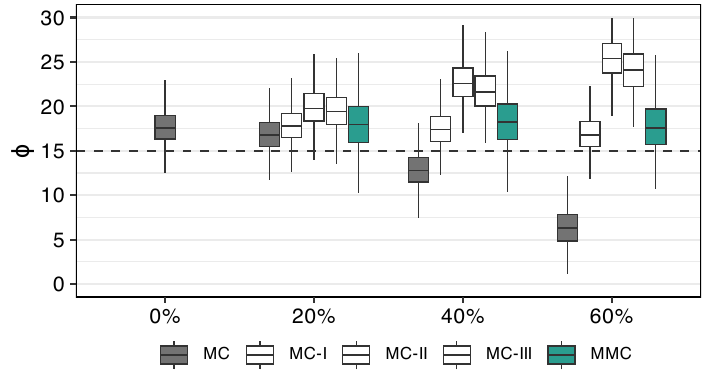}
    \end{subfigure}
    \begin{subfigure}{0.42\textwidth}
        \includegraphics[scale=0.55]{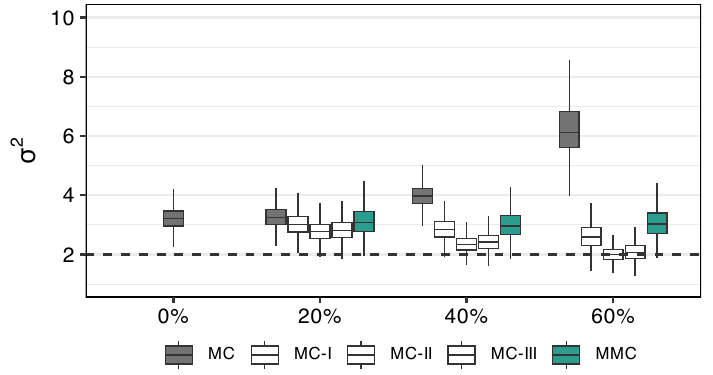}
    \end{subfigure}
    \begin{subfigure}{0.42\textwidth}
    \caption{IH2.2 (districts): $(\phi=20, \sigma^2=2)$}
        \includegraphics[scale=0.55]{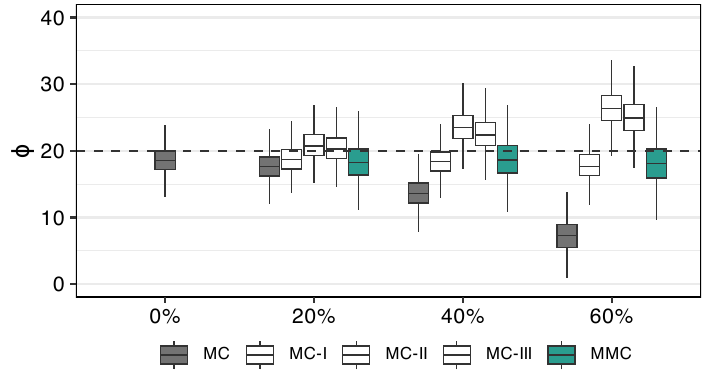}
    \end{subfigure}
    \begin{subfigure}{0.42\textwidth}
        \includegraphics[scale=0.55]{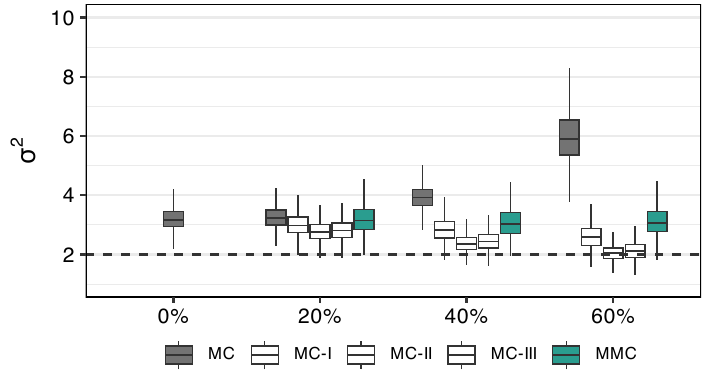}
    \end{subfigure}
    \begin{subfigure}{0.42\textwidth}
    \caption{IH2.3 (districts): $(\phi=30, \sigma^2=2)$}
        \includegraphics[scale=0.55]{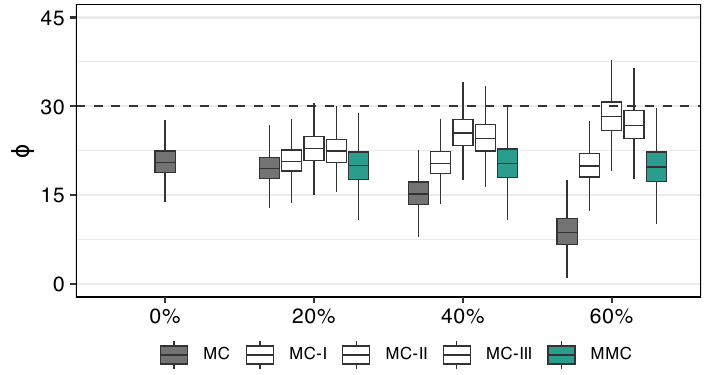}
    \end{subfigure}
    \begin{subfigure}{0.42\textwidth}
        \includegraphics[scale=0.55]{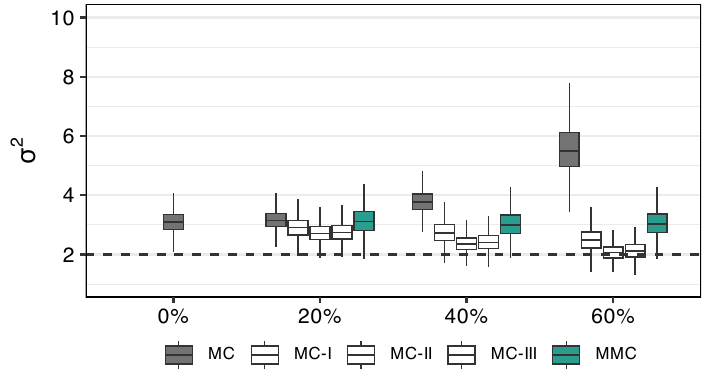}
    \end{subfigure}
    \caption{Same as Figure~\ref{fig:corruptLevels_hr}, except that it is for inhomogeneous cases (IH2) across different levels of data corruption (random).  } 
    \label{fig:corruptLevels_ih2r_cvl}
\end{figure}

%%%%%
Figure \ref{fig:corruptLevels_ih2r_cvl} presents the results for this case. 
We observe that MC estimates for data with no corruption deviate from true parameters as $\phi$ increases. 
The relative performance of the methods under these conditions is similar to the previous two cases (H and IH1). 
Here, however, the estimates of $\phi$ are not as close to the truth as the previous cases. 
This is expected given the complex mean structure of the log intensity function that may interfere with the second-order clustering structure of the point patterns. 
Indeed, even with no corruption, the estimates for $\phi$ using the MC method are somewhat far from the truth. 
It is reassuring, however, that the MMC approach gives roughly similar estimates for $\phi$ and $\sigma^2$ under all levels of corruption considered, producing values similar to those from standard MC with no data corruption. 

%%%%%%%%%%%%%%%%%%%%%%%%%%%%%%%%%%%%%%%%%%%%%%%%%%
\subsection{Selection of \texorpdfstring{$\delta$}{}}\label{subsec:tuning_pars}
%%%%%
The selection of the tuning parameter $\delta$ for the MMC method, which determines the lower bound of integration in the discrepancy measure \eqref{eqn:U_d}, plays a crucial role in mitigating the effects of duplicate points.
This section provides practical guidance for selecting an optimal $\delta$ in MMC and explores its relationship with the corruption process.
To motivate our selection criterion, we analyze the behavior of MMC estimates across a range of $\delta$ values (0 to 70) using the homogeneous case H.3. 

%%%%%
\begin{figure}[htb!]
\captionsetup[subfigure]{justification=centering}
    \centering
    \begin{subfigure}{0.24\textwidth}
    \centering
    \subcaption{No corruption}
        \includegraphics[scale=0.45]{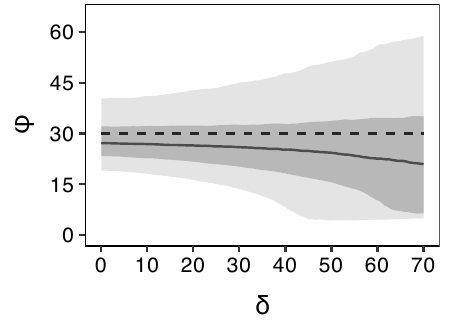}
    \end{subfigure}
    \begin{subfigure}{0.24\textwidth}
    \centering
    \subcaption{20\% corruption}
        \includegraphics[scale=0.45]{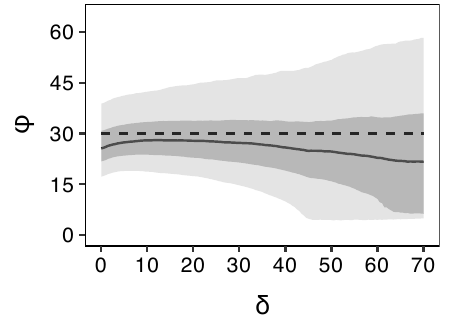}
    \end{subfigure}
    \begin{subfigure}{0.24\textwidth}
    \centering
    \subcaption{40\% corruption}
        \includegraphics[scale=0.45]{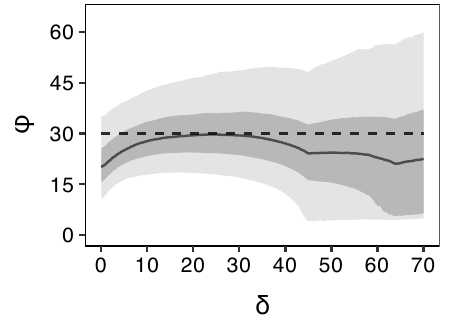}
    \end{subfigure}
    \begin{subfigure}{0.24\textwidth}
    \centering
    \subcaption{60\% corruption}
        \includegraphics[scale=0.45]{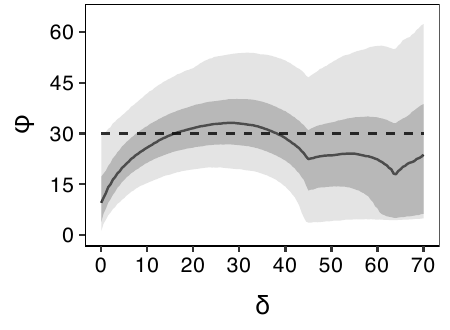}
    \end{subfigure}
    \begin{subfigure}{0.24\textwidth}
    \centering
        \includegraphics[scale=0.45]{figures/mmc_curve_h_30phi_2sig2_phi.pdf}
    \end{subfigure}
    \begin{subfigure}{0.24\textwidth}
    \centering
        \includegraphics[scale=0.45]{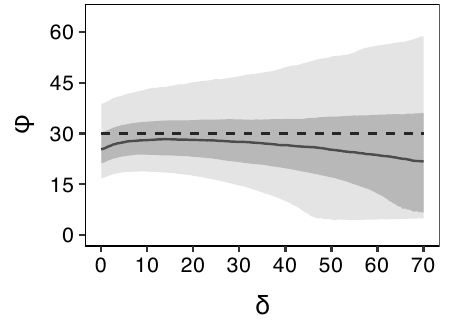}
    \end{subfigure}
    \begin{subfigure}{0.24\textwidth}
    \centering
        \includegraphics[scale=0.45]{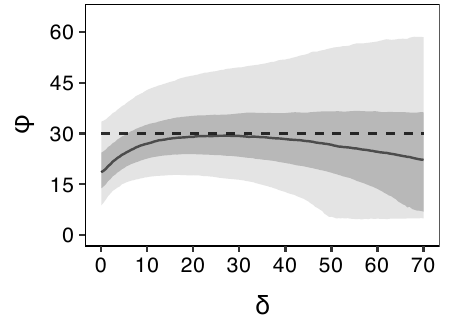}
    \end{subfigure}
    \begin{subfigure}{0.24\textwidth}
    \centering
        \includegraphics[scale=0.45]{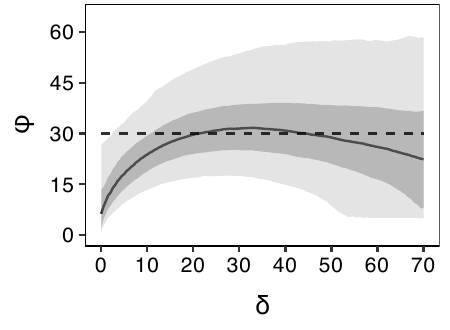}
    \end{subfigure}
    \caption{MMC estimates of $\phi$ against $\delta$ for scenario H.3 ($\phi=30$). 
    Top: corruption by the regular grid with grid cell size $45\times 45$. 
    Bottom: corruption by the tessellation with average cell area $45^2$. 
    The solid lines  show the median of the MMC estimates.
    The dark grey regions give the interquartile range, and the light grey givesbounds between the 5th and 95th percentiles. 
    The dashed lines mark the true parameter value.
    } \label{fig:mmc_curve_h_30phi}
\end{figure}

%%%%%
Figure \ref{fig:mmc_curve_h_30phi} illustrates how $\delta$ influences the estimated spatial range parameter $\phi$ under different levels of corruption.
When no corruption is present (column \paren{a}), the MMC estimates of $\phi$ remain mostly stable across all $\delta$ values. 
This suggests that in the absence of duplicates, MMC is not highly sensitive to this parameter, at least up to a certain point. 
However, as the corruption level increases, the MMC curves exhibit concave patterns, particularly when data are corrupted on a regular grid.
Notably, as shown in the top row of Figure \ref{fig:mmc_curve_h_30phi}, sharp changes in MMC estimates occur around $\delta{\,\approx\,}45$ and $\delta{\,\approx\,}63.6$, corresponding to the side length and diagonal length of the grid cell, respectively. 
These artifacts disappear when data corruption is applied using tessellations (bottom row of the figure). 
This reinforces the notion that $\delta$ should be selected based on the geometry of the data corruption process (i.e., the shape and size of domain partitions) rather than the underlying second-order properties of the point patterns.  

%%%%%
% Here we suggest using the average of the distance that a point is snapped to the centroid of the corresponding partition as the optimal $\delta$. 
Based on these observations, we propose to set $\delta{\,=\,}\frac{1}{3}D$, where $D$ is the diameter of a circle with the equal area of the grid cell or the district. 
% If there are districts of different sizes, we propose to use one-third of the average of the maximum distances for all the districts. 
This rule reflects the expected displacement of points that have been randomly snapped to centroids within their partitions.
For example, in a $[0,810]{\,\times\,}[0,810]$ square domain partitioned into an 18 by 18 regular grid or a tessellation with 324 polygons, the average cell area is $45^2$. 
Applying our rule, the corresponding optimal $\delta$ value is $\frac{1}{3}\left(2\sqrt{45^2/\pi}\right)\approx17$. 
This approach is especially useful when researchers possess prior information on the corruption process, as is often the case for social and political events data -- e.g., knowledge of snapping locations in the crime data or the ``specificity'' item in the GTD data set discussed in Section \ref{sec:motivation}.

\begin{figure}[htb!]
    \centering
    \begin{subfigure}{0.4\textwidth}
    \caption{H.1 (regular grid): $(\phi=15, \sigma^2=2)$}
        \includegraphics[scale=0.5]{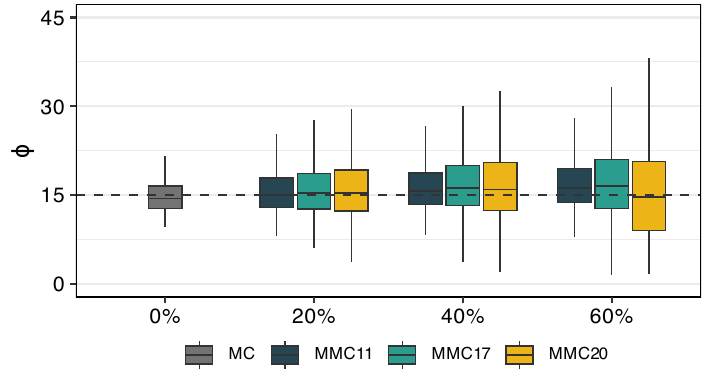}
    \end{subfigure}
    \begin{subfigure}{0.4\textwidth}
        \includegraphics[scale=0.5]{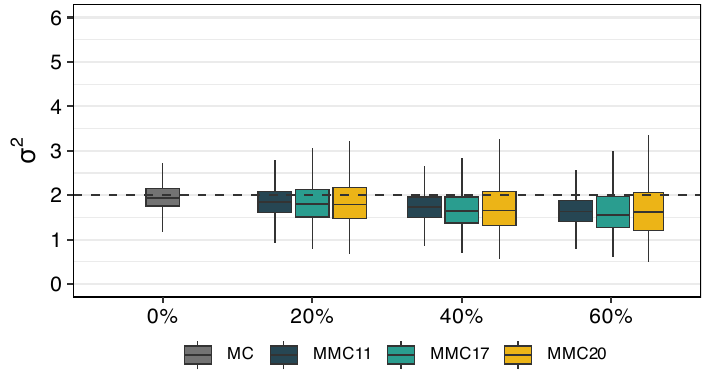}
    \end{subfigure}
    \begin{subfigure}{0.4\textwidth}
    \caption{H.2 (regular grid): $(\phi=20, \sigma^2=2)$}
        \includegraphics[scale=0.5]{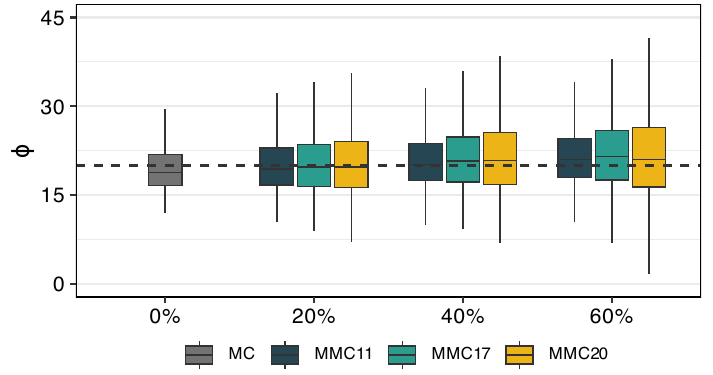}
    \end{subfigure}
    \begin{subfigure}{0.4\textwidth}
        \includegraphics[scale=0.5]{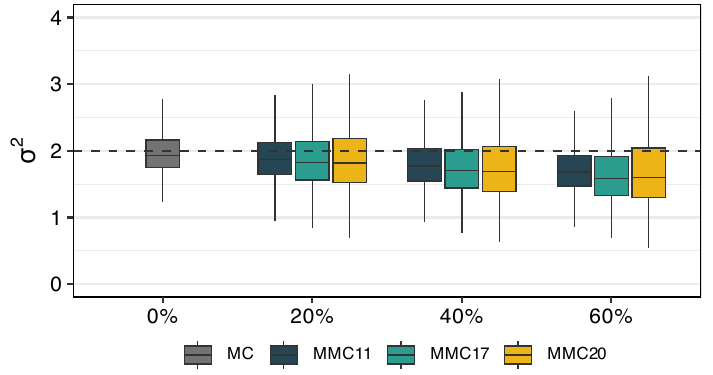}
    \end{subfigure}
    \begin{subfigure}{0.4\textwidth}
    \caption{H.3 (regular grid): $(\phi=30, \sigma^2=2)$}
        \includegraphics[scale=0.5]{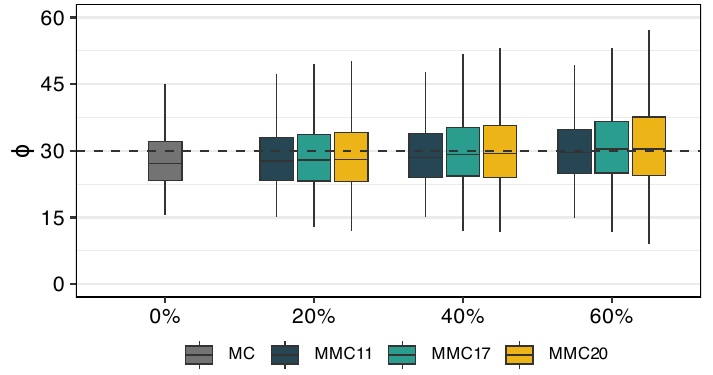}
    \end{subfigure}
    \begin{subfigure}{0.4\textwidth}
        \includegraphics[scale=0.5]{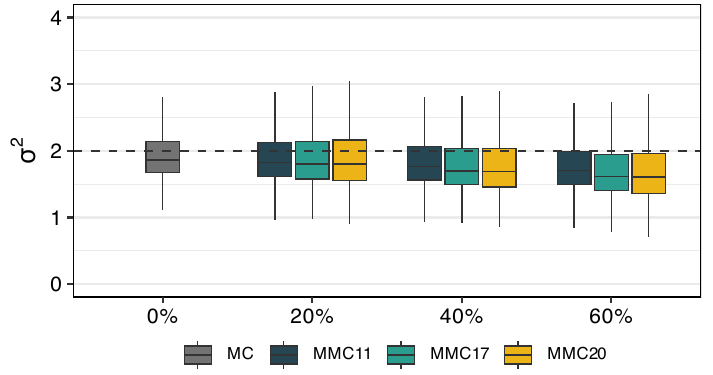}
    \end{subfigure}
    \caption{Estimates of $\phi$ and $\sigma^2$ for H.1-H.3 cases across various corruption levels on a regular grid. Data were corrupted at three spatial resolutions with cell areas $30^2$, $45^2$, and $54^2$. 
    MMC estimates are shown for different $\delta$ values (11, 17, and 20). 
    The dashed lines represent the true parameter values. } 
    \label{fig:mmc_delta_grid}
\end{figure}

\begin{figure}[htb!]
    \centering
    \begin{subfigure}{0.4\textwidth}
    \caption{H.1 (tessellation): $(\phi=15, \sigma^2=2)$}
        \includegraphics[scale=0.5]{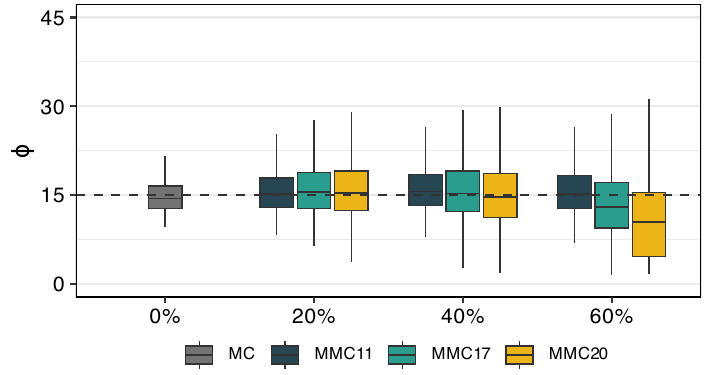}
    \end{subfigure}
    \begin{subfigure}{0.4\textwidth}
        \includegraphics[scale=0.5]{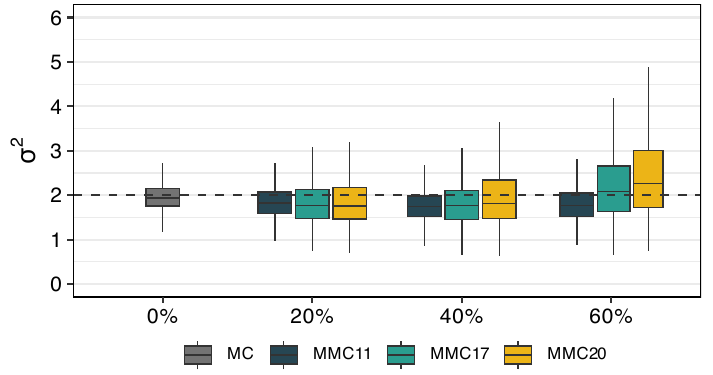}
    \end{subfigure}
    \begin{subfigure}{0.4\textwidth}
    \caption{H.2 (tessellation): $(\phi=20, \sigma^2=2)$}
        \includegraphics[scale=0.5]{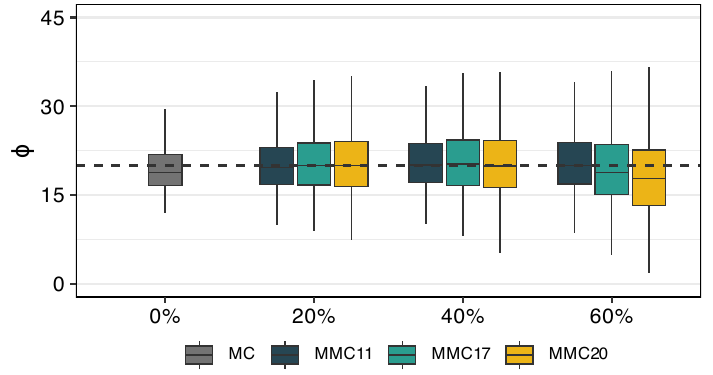}
    \end{subfigure}
    \begin{subfigure}{0.4\textwidth}
        \includegraphics[scale=0.5]{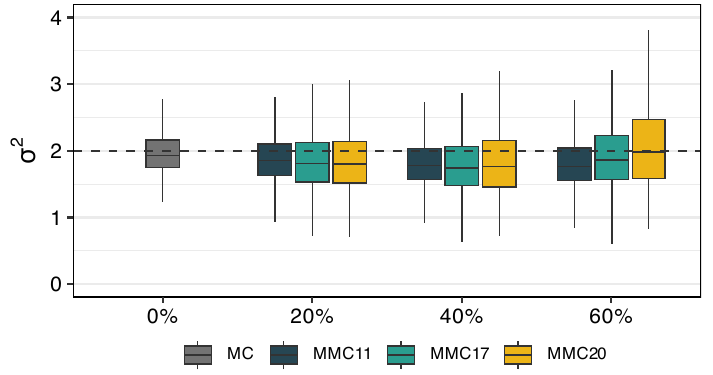}
    \end{subfigure}
    \begin{subfigure}{0.4\textwidth}
    \caption{H.3 (tessellation): $(\phi=30, \sigma^2=2)$}
        \includegraphics[scale=0.5]{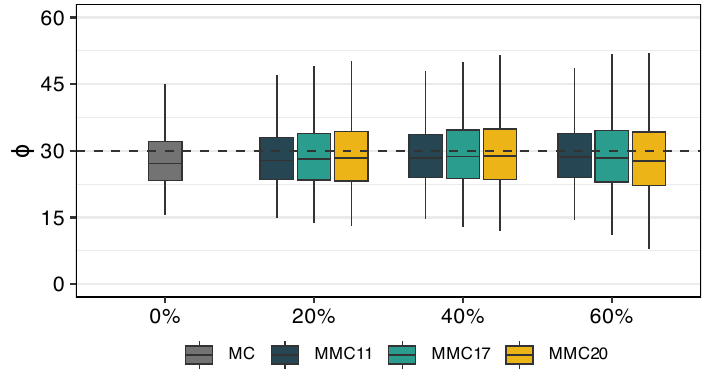}
    \end{subfigure}
    \begin{subfigure}{0.4\textwidth}
        \includegraphics[scale=0.5]{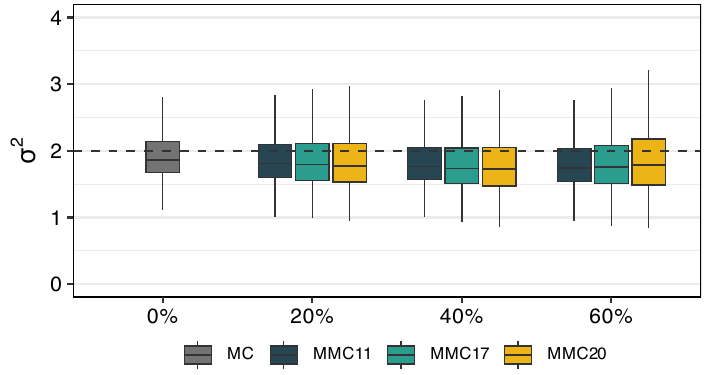}
    \end{subfigure}
    \caption{Same as Figure~\ref{fig:mmc_delta_grid}, except that corruptions are based on a tessellation grid.} 
    \label{fig:mmc_delta_tess}
\end{figure}

%%%%%
To validate this selection criterion, we apply MMC to homogeneous cases (H.1-H.3) with corruption applied at three spatial resolutions: 729, 324, and 225 partitions, yielding average cell areas of $30^2$, $45^2$, and $54^2$, respectively.
According to our rule, the corresponding $\delta$ values are 11, 17, and 20.
Figures \ref{fig:mmc_delta_grid} and \ref{fig:mmc_delta_tess} illustrate the sensitivity of these choices on parameter estimation.
For corruption on a regular grid (Figure \ref{fig:mmc_delta_grid}), the one-third rule performs well across different resolutions and corruption levels. 
Similar findings are observed for tessellation-based corruption (Figure \ref{fig:mmc_delta_tess}). 
With higher spatial resolution (i.e., smaller cell areas $30^2$), the one-third rule works well for all homogeneous cases.
However,  for coarser spatial resolutions (average cell areas $45^2$ and $54^2$), MMC estimates deviate from the true parameters at high corruption levels (60\%) for cases H.1 and H.2. 
In contrast, for H.3, where spatial dependence is stronger ($\phi{\,=\,}30$), MMC estimates remain stable across all resolutions.
Overall, these findings suggest that the ``rule of thirds'' provides a practical and reliable heuristic for selecting $\delta$.

%%%%% (H) fig: MMC with delta = 1/3 grid cell size (regular grid) 

%%%%% (H) fig: MMC with delta = 1/3 grid cell size (tessellation) 

\section{Application}\label{sec:application}
\begin{figure}[htb!]
\centering
    \includegraphics[scale=1]{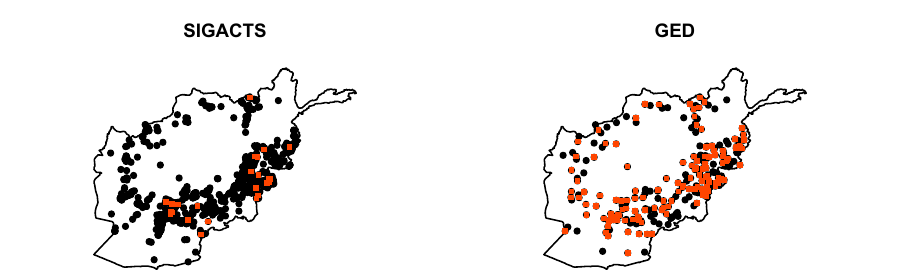}
    \caption{Point patterns of SIGACTS and GED matched data. Black dots represent unique points, and orange duplicated points.}
    \label{fig:sppatterns_paired}
\end{figure}

%%%%%
To demonstrate the efficacy of our proposed methods on real-world data, we now revisit one of our motivating examples: Conflict events in Afghanistan from 2008-2009, the SIGACTS and GED data. 
Figure~\ref{fig:sppatterns_paired} shows spatial point patterns from the SIGACTS and GED data sets, including spatial locations with duplicated data. 
We consider an inhomogeneous LGCP model for both datasets. 
We use the results from SIGACTS as the reference case since, as we discussed above, the location coordinates in SIGACTS are considered more accurate than the GED data; out of the 1,077 matched events, 26 duplicated points were found in the SIGACTS data, while the GED data contained 826 duplicated points.
The adaptive kernel approach is used to estimate the intensity for both SIGACTS and GED data, with the global bandwidth $h_0{\,=\,}100$ (see equation \eqref{eqn:inhom_lamEst2} in the Appendix).
% For the bandwidth for the kernel, we set $h{\,=\,}100$. 

%%%%%

%%%%%
\begin{figure}[htb!]
    % \centering
    \begin{subfigure}[b]{\textwidth}
        \includegraphics[scale=0.71]{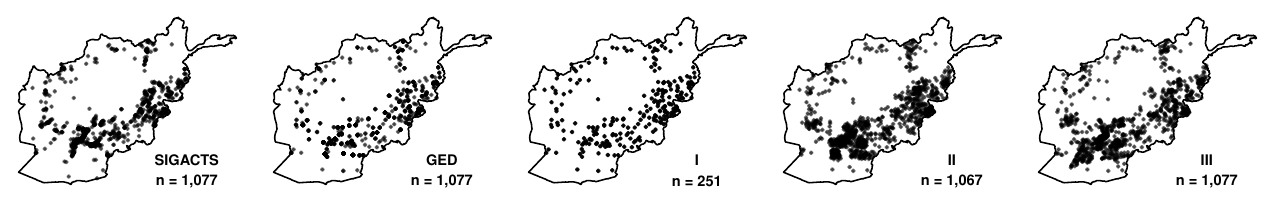} 
    \end{subfigure}
    \vfill
    \begin{subfigure}[b]{\textwidth}
        \includegraphics[scale=0.76]{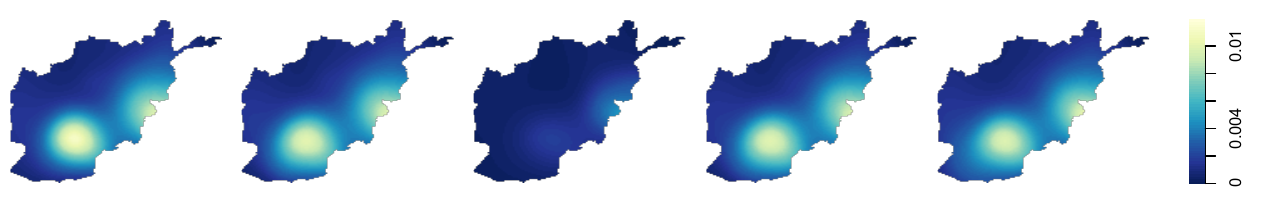}
    \end{subfigure}
    \vfill
    \caption{Top: SIGACTS, GED and GED with Methods \Romannum{1}, \Romannum{2}, and \Romannum{3} to deal with duplicates, respectively. 
    Bottom: corresponding estimated intensity by adaptive kernel approach ($h_0{\,=\,}100$).  
    }
    \label{fig:paired_inten}
\end{figure}

%%%%%
Figure \ref{fig:paired_inten} presents the estimated first-order intensity surfaces for SIGACTS, GED, and GED with three methods to handle the duplicates. 
Other than Method \Romannum{1}, all other cases for SIGACTS and GED result in comparable first-order intensity surfaces.  
Overall, we observe two high-intensity spots for the estimated intensity from SIGACTS, one in the east over the capital city (Kabul) and the other in the south of the country near Kandahar city. 
As expected, due to the deletion of events, using Method \Romannum{1}, the level of estimated first-order intensity is much lower than the remaining cases, which may lead to incorrect inferences.

%%%%%
Table \ref{tab:appl_compare_methods} summarizes the estimated spatial range parameter ($\phi$) and variance ($\sigma^2$) for each method. 
Based on our one-third rule in Section \ref{subsec:tuning_pars} for choosing $\delta$, we fix $\delta{\,=\,}17$ in MMC for both SIGACTS and GED datasets.
For the SIGACTS data, all methods yield similar estimates of $\phi$ (approximately 18-22 km) and $\sigma^2$ (ranging from 3.09 to 3.51). 
This is expected given a small percentage of duplicated points in the SIGACTS data. 
Note that the MMC method provides $\phi$ and $\sigma^2$ estimates comparable to all other methods. 
The estimated $\phi$ value may seem somewhat large, but the deviation is within the variation of estimates in a comparable simulation case (IH2.2) in Figure~\ref{fig:corruptLevels_ih2r_cvl}. 
A reason why the estimate of $\phi$ for the MMC method is slightly larger may be the truncation of $K$-function values up to relatively large $\delta$, despite the fact that there are only few duplicates for SIGACTS data. 
As a result, the variation of the estimates for MMC may be larger than that for MC estimates in this case.
%It may be possible to assess the uncertainty of these estimates using a bootstrap approach. 

%%%%%
\begin{table}[htb!]
\captionsetup{width=5in}
%% h0 = 100
\caption{Comparison of parameter estimates ($\phi,\sigma^2$)  for SIGACTS and GED data. 
% Global bandwidth $h=100$.
The jittering radius in MC-\Romannum{2} is 25 (km).
$\delta$ in MMC is 17 (km). 
}\label{tab:appl_compare_methods}
\centering
\adjustbox{width=0.45\textwidth}{
\begin{tabular}{lcccccc}
\toprule
 & \multicolumn{2}{c}{\textbf{SIGACTS}} & \multicolumn{2}{c}{\textbf{GED}}\\
\cmidrule(lr){2-3}\cmidrule(lr){4-5}
\textbf{Method} & $\phi$ (km) & $\sigma^2$ & $\phi$ (km) & $\sigma^2$ \\
\midrule
\textbf{MC}      & 18.52 & 3.50 &  0.71 & 12.47 \\
\textbf{MC-I}    & 18.33 & 3.51 & 50.00 &  0.43 \\
\textbf{MC-II}   & 18.46 & 3.49 & 29.95 &  1.72 \\
\textbf{MC-III}  & 18.69 & 3.47 & 23.38 &  2.15 \\
\textbf{MMC}     & 21.55 & 3.09 & 17.52 &  3.21 \\ 
\bottomrule                      
\end{tabular}
}
\end{table}

%%%%%
On the other hand, the GED data reveal stark differences across methods.
The standard MC method, which does not account for duplicates, produces an implausibly small $\phi$ (0.71 km) and an inflated variance estimate ($\hat{\sigma}^2{\,=\,}12.47$).
MC-I gives much larger estimate for $\phi$ (50 km) compared to those values for SIGACTS data and thus substantially smaller value of estimate for $\sigma^2$ (0.43).
MC-II and MC-III also show some level of overestimation for $\phi$ and underestimation for $\sigma^2$ (compared to the estimates for the SIGACTS data).
In contrast, the MMC method provides estimates that closely align with the MC estimates from SIGACTS, which demonstrates the utility of the proposed method.

\section{Discussion}\label{sec:discussion}
%%%%% conclusion and key findings
In this study, we consider the challenges of analyzing and modeling spatial point patterns with duplicate points. 
Existing approaches to dealing with duplicate data often involve altering the original data, either removing duplicate observations or adding noise to reported locations. While these strategies may be fine if the number of duplicate observations is low, real-world data is often replete with duplicate observations (over 80\% of the sample in some examples discussed here) due to recording imprecision or snapping. In these instances, we advocate an approach that modifies the method of inference to address duplicates in estimation, thereby reducing the need for researchers to make \emph{ad hoc} distortions to the original data.

%%%%%
Specifically, we demonstrate how the MC method can be modified to account for duplicates by truncating the lower bound of integration in the discrepancy measure with a tuning parameter, $\delta$. 
In doing so, only non-duplicate discrepancies are included in estimation of the second-order parameters. Using simulation experiments, we show that when duplicates result from the geo-coding process, a natural choice for $\delta$ emerges -- i.e., the ``rule of thirds'' -- to help guide the proposed MMC method. 
%Specifically, the optimal $\delta$ is a more a function of the data corruption process (e.g. the extent of the geo-location error) than the underlying spatial point pattern structure. With some prior knowledge of this process, the selection   
In both simulated and real-data applications, the MMC method outperforms the general MC methods with existing approaches in handling duplicates. 
Where previous methods sought to obscure (jitter) or disregard (delete), our proposed method directly addresses the problem created by duplicates -- the erroneous discrepancy at the micro-scale -- by identifying where to begin measuring the true discrepancy. These performance differences are especially pronounced when the number of duplicates is large, as with the GED data in Section~\ref{sec:application}. 

%%%%%
There are several ways that our findings can be extended in future research. First, we illustrated MMC using the LGCP model, however, the suggested modification is quite general. Any parametric point process with an explicit form of the $K$-function (e.g., Neyman–Scott or Thomas processes) can similarly incorporate a positive lower bound in the minimum contrast criterion. Second, while past work has discussed the use of non-zero lower bounds in  MC estimation, which has typically focused on numerical instability when using the $g$-function, our results suggest researchers may want to consider other contexts in which a positive lower bound could prove advantageous. Third, in our application, we report only the point estimates of the parameters without accompanying uncertainty bounds (e.g., standard errors). Although several works \citep{Heinrich1992, Guan2007, Zhu2025} have studied the asymptotic properties of MC estimators and derived asymptotic variance expressions, these results do not directly extend to our MMC framework or to data subject to duplication.
Further research is needed to establish theoretical variance in this setting.

%%%%% references
% \clearpage
\bibliographystyle{styles/asa}
\bibliography{references}

%%%%% appendices
\clearpage
% \pagenumbering{arabic}
\renewcommand{\thesection}{\Alph{section}.\arabic{section}}
\setcounter{section}{0}
\counterwithin{figure}{section}
\numberwithin{equation}{section}

\appendix
\noindent {\Large \bf Appendix }
\section{First-order Intensity Estimation and Bandwidth Selection} 
%%%%%
The estimation of the empirical $K$-function, whether $\widehat{K}_{\thom}$ in equation \eqref{eqn:estKhom} or $\widehat{K}_{\tinhom}$ in equation \eqref{eqn:estKinhom}, requires evaluating the first-order intensity function $\lambda$. 
Below, we detail the methods for estimating $\lambda$ and the critical considerations for bandwidth selection in both homogeneous and inhomogeneous cases.

%%%%% Homogeneous case (H)
For homogeneous point patterns (scenario H), the intensity $\lambda$ is constant, representing the expected number of points per unit area.
We estimate it as $\lamhat {\,=\,} \frac{N-1}{|W|}$, where $N$ is the number of observed points and $|W|$ is the area of the study window.
This estimation aligns with the implementation in the \textit{Kest} function from the R package \textit{spatstat} \citep{Baddeley2015}.

%%%%% Inhomogeneous case (IH1)
For inhomogeneous cases, we estimate the first-order intensity nonparametrically. 
Specifically, for the IH1 scenario, which features a smoothly varying intensity structure, we use a classic fixed-bandwidth kernel density estimator with a Gaussian kernel \citep{Silverman2018, Davies2018a}. 
Formally, for $n$ observed points $\bunderline{X}{\,=\,}\{\bx_i\}_{i=1}^n$, the fixed-bandwidth kernel estimate of $\lambda$ is written as 
\begin{align}\label{eqn:inhom_lamEst1}
    \hat{\lambda}_h(\bs\,|\,\bunderline{X}) &= n^{-1}h^{-2}\,\sumin\, \kappa\Big(\frac{\bs-\bx_i}{h}\Big)\,q_h(\bx_i\,|\,W)^{-1}, \qquad \bs\in W,\\[2pt] \nonumber 
    q_h(\bx_i\,|\,W) &= h^{-2}\,\int_W \kappa\Big( \frac{\bu-\bx_i}{h} \Big)\,\td\bu, 
\end{align}
where $h{\,>\,}0$ is the bandwidth, $\kappa(\cdot)$ is the kernel function and $q_h(\cdot)$ is an edge-correction factor.
This method provides a straightforward way to capture spatial trends but requires careful selection of the bandwidth parameter $h$.

%%%%% Inhomogeneous case (IH2)
On the other hand, for the IH2 scenario involving a more complex intensity structure, we adopt the adaptive kernel approach \citep{Davies2018, Davies2018a}.
This method allows the bandwidth to vary spatially: in areas of high density, the resulting bandwidth is small, while in areas of low density, the resulting bandwidth is large. 
Formally, the adaptive estimator is given as
\begin{align}\label{eqn:inhom_lamEst2}
    \Tilde{\lambda}_{h_0}(\bs\,|\,\bunderline{X}) &= n^{-1}\,\sumin\, h(\bx_i\,;\lambda)^{-2} \,\kappa\Big( \frac{\bs-\bx_i}{h(\bx_i\,;\lambda)} \Big)\,q_{h{(\bx_i;\lambda)}}(\bx_i\,|\,W)^{-1},\\[2pt] \nonumber 
    h(\bu;\lambda) &= h_0\,\hat{\lambda}_{\hat{h}}(\bu\,|\,\bunderline{X})^{-\frac{1}{2}}\,\gamma_\lambda^{-1}, \quad \bs\in W, 
\end{align}
where the smoothing bandwidth $h(\cdot)$ is a function of a coordinate, $\hat{\lambda}$ is a pilot estimate of the unknown intensity constructed via (\ref{eqn:inhom_lamEst1}) with a fixed pilot bandwidth $\hat{h}$, $h_0$ is the global bandwidth which controls the global smoothing of the variable bandwidths, $\gamma_\lambda$ is the is the geometric mean of the inverse-density bandwidth factors, and $q_{h{(\bx_i;\lambda)}}$ is the edge correction for the adaptive estimator.
For greater details of this approach, we refer to \cite{Abramson1982}, \cite{Silverman1986}, and \cite{Davies2018a}.
The adaptive kernel approach is implemented using the function \textit{bivariate.density} in the \textit{sparr} {\sf R} package, with the global bandwidth $h_0$ (a scalar value) needing to be predetermined. 

%%%%%
The choice of bandwidth -- whether the fixed-bandwidth $h$ for classic kernel smoothing or the global bandwidth $h_0$ for adaptive kernel smoothing -- is a pivotal and challenging task \citep{CvL2018}.
% which has an obvious impact on the resulting minimum contrast estimates of the second-order parameters. 
Bandwidth selection directly impacts the accuracy of the second-order parameter estimates derived from the minimum contrast method.
As noted by \citet{Davies2013}, if the bandwidth is too large, we overestimate the true level of dependence, while too small a bandwidth leads to missing any correlation whatsoever.

%%%%% bandwidth selectors and RED
There are a number of methods in the literature for selecting bandwidth: four approaches (\textit{CvL}, \textit{diggle}, \textit{ppl}, \textit{scott}) are available in the \textit{spatstat} {\sf R} package \citep{Baddeley2015} and another four approaches (\textit{BOOT}, \textit{LIK}, \textit{LSCV}, \textit{OS}) available in the \textit{sparr} {\sf R} package \citep{Davies2018a}. 
We refer to \cite{Baddeley2015} and \cite{Davies2018a} for details of each method. 
We tested all of these methods for the two inhomogeneous simulation scenarios (IH1 and IH2).
For comparison, we also employed the Relative Euclidean Distance (RED) criterion used in \citet{Davies2013}, which elects the bandwidth that minimizes the deviation of estimated second-order parameters (i.e., $\phi$ and $\sigma^2$) from their true values for uncorrupted data. 
% For the two inhomogeneous simulation scenarios (IH1 and IH2), we evaluated these methods alongside a modified \textit{Relative Euclidean Distance} (RED) criterion \citep{Davies2013}, which elects the bandwidth that minimizes the deviation of estimated second-order parameters (i.e., $\phi$ and $\sigma^2$) from their true values for uncorrupted data.
The \textit{RED} is defined as 
\begin{equation*}
    \text{RED}(\bar{\phi},\bar{\sigma}^2){\,=\,}\sqrt{\text{ARE}(\bar{\phi})^2+\text{ARE}(\bar{\sigma}^2)^2},
\end{equation*}
where $\text{ARE}(\bar{\phi})=|\bar{\phi}-\phi|/\phi$ and $\text{ARE}(\bar{\sigma}^2){\,=\,}|\bar{\sigma}^2-\sigma^2|/\sigma^2$ are \textit{absolute relative errors} of MC estimates ($\bar{\phi}$ and $\bar{\sigma}^2$) from uncorrupted data.  
The selected bandwidth corresponds to the one that minimizes \textit{RED}.

\begin{figure}[H]
\captionsetup[subfigure]{justification=centering}
    \centering
    \begin{subfigure}{0.32\textwidth}
    \centering
    \subcaption{IH1.1 ($\phi=15, \sigma^2=2$)}
        \includegraphics[scale=0.52]{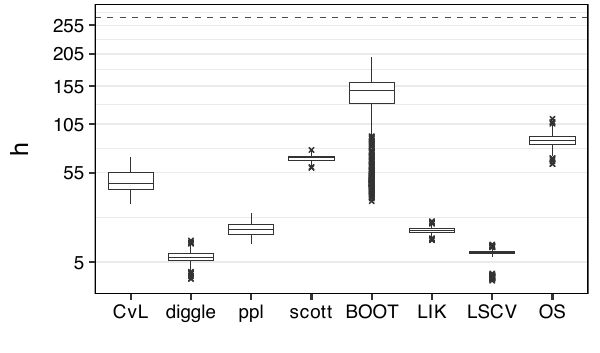}
    \end{subfigure}
    \begin{subfigure}{0.32\textwidth}
    \centering
    \subcaption{IH1.2 ($\phi=20, \sigma^2=2$)}
        \includegraphics[scale=0.52]{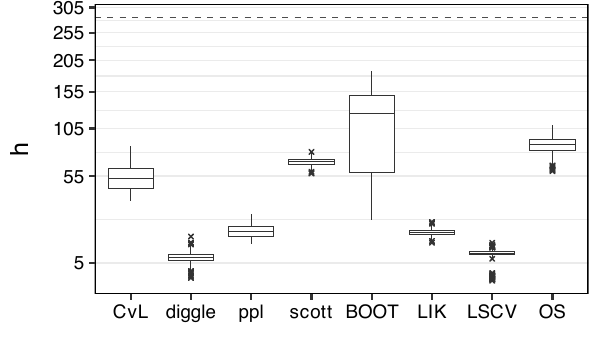}
    \end{subfigure}
    \begin{subfigure}{0.32\textwidth}
    \centering
    \subcaption{IH1.3 ($\phi=30, \sigma^2=2$)}
        \includegraphics[scale=0.52]{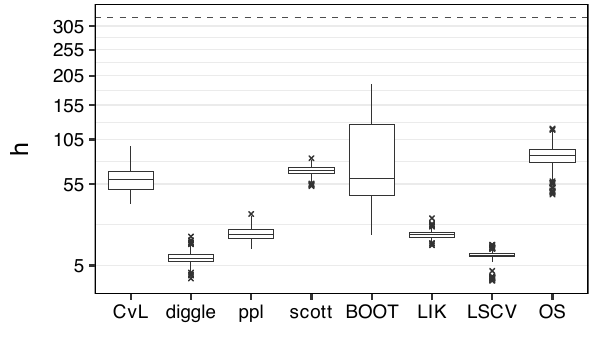}
    \end{subfigure}
    \caption{Bandwidth $h$ for IH1 cases selected by eight different approaches, where the dashed line represents the one chosen by the criterion based on RED. Each boxplot is based on 1,000 simulations. }
    \label{fig:ih1_bw_selectors}
\end{figure}

\begin{figure}[H]
\captionsetup[subfigure]{justification=centering}
    \centering
    \begin{subfigure}{0.32\textwidth}
    \centering
    \subcaption{IH2.1 ($\phi=15, \sigma^2=2$)}
        \includegraphics[scale=0.52]{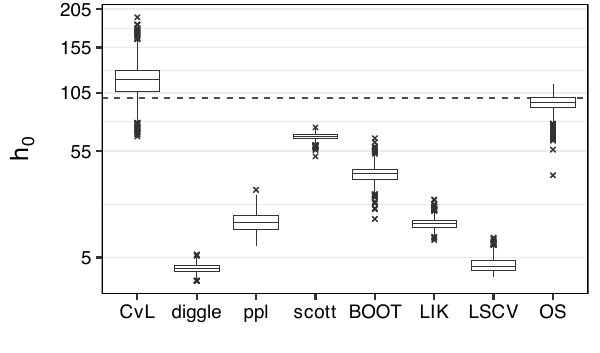}
    \end{subfigure}
    \begin{subfigure}{0.32\textwidth}
    \centering
    \subcaption{IH2.2 ($\phi=20, \sigma^2=2$)}
        \includegraphics[scale=0.52]{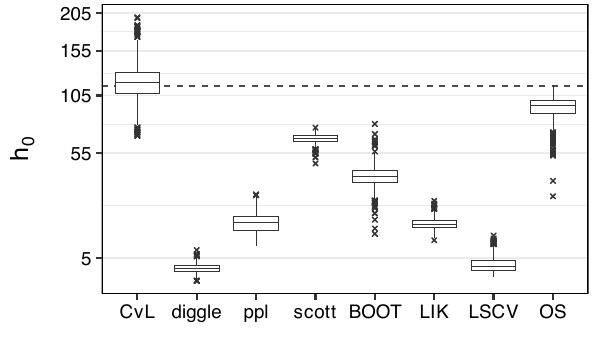}
    \end{subfigure}
    \begin{subfigure}{0.32\textwidth}
    \centering
    \subcaption{IH2.3 ($\phi=30, \sigma^2=2$)}
        \includegraphics[scale=0.52]{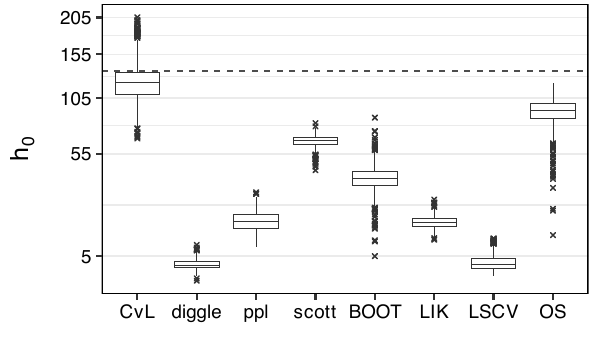}
    \end{subfigure}
    \caption{Same as Figure~\ref{fig:ih1_bw_selectors}, except that this is for global bandwidth $h_0$ for IH2 cases. }
    \label{fig:ih2_bw_selectors}
\end{figure}

%%%%%
We found that, for IH1, all of these methods give values of bandwidth that are far too small (as seen in Figure \ref{fig:ih1_bw_selectors}), resulting in a dramatic underestimation of the true level of dependence.
Hence, we use bandwidths selected by the \textit{RED} criterion in this case.
On the other hand, for IH2, one of the eight methods, namely $CvL$, works notably well compared to the rest of the methods (as seen in Figure \ref{fig:ih2_bw_selectors}). 
Since IH2 is the scenario with a more realistic first-order intensity structure, we use the $CvL$ method for the simulation study for the IH2 case as well as real data analysis. 

%%%%% bandwidth selectors for IH1

%%%%% bandwidth selectors for IH2

\end{document}